\documentclass[aps,twocolumn,preprintnumbers,eqsecnum,prb]{revtex4}
\usepackage{amsfonts}
\usepackage[dvips]{graphicx}
\usepackage[T1]{fontenc}
\usepackage{amsmath,amsbsy,amssymb,graphics}

\begin{document}
\preprint{Accepted for Publication in PRD}

\title{Topological Solitons in Noncommutative Plane and Quantum Hall
Skyrmions}
\author{Z.F. Ezawa$^{1}$ and G. Tsitsishvili$^{1,2}$}
\affiliation{{}$^1$Department of Physics, Tohoku University, Sendai, 980-8578 Japan\\
{}$^2$Department of Theoretical Physics, A. Razmadze Mathematical Institute,
Tbilisi, 380093 Georgia}
\date{\today}

\begin{abstract}
We analyze topological solitons in the noncommutative plane by taking a
concrete instance of the quantum Hall system with the SU(N) symmetry, where
a soliton is identified with a skyrmion. It is shown that a topological
soliton induces an excitation of the electron number density from the
ground-state value around it. When a judicious choice of the topological
charge density $J_{0}(\mathbf{x})$ is made, it acquires a physical reality
as the electron density excitation $\Delta \rho ^{\text{cl}}(\mathbf{x})$
around a topological soliton, $\Delta \rho ^{\text{cl}}(\mathbf{x})=-J_{0}(%
\mathbf{x})$. Hence a noncommutative soliton carries necessarily the
electric charge proportional to its topological charge. A field-theoretical
state is constructed for a soliton state irrespectively of the Hamiltonian.
In general it involves an infinitely many parameters. They are fixed by
minimizing its energy once the Hamiltonian is chosen. We study explicitly
the cases where the system is governed by the hard-core interaction and by
the noncommutative CP$^{N-1}$ model, where all these parameters are
determined analytically and the soliton excitation energy is obtained.
\end{abstract}

\maketitle

\section{Introduction}

Noncommutative geometry\cite{BookConnes} has attracted much attention in a
recent development of string theory, as triggered in part by the invention
of noncommutative instantons\cite{Nekrasov98} and solitons\cite{Gopakumar00}%
. Since then, noncommutative solitons have been studied in various
noncommutative field theories such as the Abelian Higgs model\cite%
{Schaposnik04} and the CP$^{N-1}$ model\cite{LeePLB01}.

The standard way of constructing a noncommutative field theory is to replace
the ordinary product by the Groenewold-Moyal\cite{groenewold,moyal} product (%
$\star $-product) in the Lagrangian density of the corresponding commutative
theory. The replacement leads to interesting features reflecting the
peculiarities of noncommutativity. In this light the noncommutative field
theory appears as a theory of classical fields and comprise no underlyeing
microscopic (second quantized) origin.

On the other hand, it has long been known that the quantum Hall (QH) system%
\cite{BookEzawa,BookDasSarma} has its essence in the noncommutativity\cite%
{Girvin84B,Girvin85L} of the coordinate of an electron in the lowest Landau
level and that its algebraic structure\cite%
{Iso92PLB,Cappelli93NPB,Moon95B,Ezawa97B} is W$_{\infty }$. A charged
excitation is a topological soliton identified with a skyrmion\cite%
{Sondhi93B}, whose existence has been confirmed experimentally\cite%
{Barrett95L,Schmeller95L,Aifer96L} by measuring the number of flipped spins
per excitation. Accordingly there must be a microscopic soliton state in a
consistent quantum field theory on the noncommutative plane. Indeed, a
candidate of the microscopic skyrmion state was proposed\cite%
{Fertig94B,MacDonald96B,Abolfath97B} to carry out a Hartree-Fock
approximation to estimate its excitation energy. Thus the QH system is an
ideal laboratory to play with noncommutative geometry and noncommutative
solitons. However, no indication of noncommutativity has so far been
reproduced in the classical picture corresponding to these microscopic
approaches.

The aim of this paper is to fill out such a lack of interplay between the
noncommutativity in the microscopic theory and the noncommutativity in the
classical theory.

Taking a concrete instance of the QH system, we investigate topological
solitons in the noncommutative plane with the coordinate $\mathbf{x}=(x,y)$.
The noncommutativity is represented in terms of the Moyal bracket as 
\begin{equation}
\lbrack x,y]_{\star }=-i\theta
\end{equation}%
with $\theta >0$. Employing the picture of planar electrons performing
cyclotron motion in strong magnetic field, we demonstrate that the
noncommutativity encoded at \textit{microscopic} level develops into the
noncommutative \textit{kinematical }properties of the corresponding \textit{%
macroscopic} objects, which are the expectation values of quantum operators.
Such an observation is quite reasonable, since the noncommutativity appears
as a property of the plane itself, while the \textit{dynamics} is the
subsequent structure built over the plane. This approach is particularly
useful to explore topological solitons since the topological charge is
essentially a geometrical property.

In the noncommutative field theory the basic object is the Weyl-ordered
operator\cite{Weyl} together with its symbol. We denote the holomorphic
basis by $\{|n\rangle ;n=0,1,\cdots \}$, in which the Weyl-ordered operator
act. We consider the planar electron field $\psi _{\mu }(\mathbf{x})$
carrying the SU(N) isospin index $\mu $ in the lowest Landau level, where
the physical variable is the U(N) density $D_{\mu \nu }(\mathbf{x})\equiv
\psi _{\nu }^{\dag }(\mathbf{x})\psi _{\mu }(\mathbf{x})$ comprising the
electron density $\rho (\mathbf{x})$ and the SU(N) isospin density $S_{a}(%
\mathbf{x})$. We define the associated bare density $\hat{D}_{\mu \nu }(%
\mathbf{x})$, subject to the so-called W$_{\infty }$(N) algebra\cite%
{EzawaX03B}, and its classical field $\hat{D}_{\mu \nu }^{\text{cl}}(\mathbf{%
x})$ that is the expectation value by the Fock state in problem. We explore
the kinematical and dynamical properties of the classical field $\hat{D}%
_{\mu \nu }^{\text{cl}}(\mathbf{x})$.

We summarize our new results. First of all, we show that the classical field
configuration, in particular the skyrmion configuration, satisfies the
noncommutative constraint 
\begin{equation}
\sum_{\sigma =1}^{N}{\hat{D}}_{\mu \sigma }^{\text{cl}}(\mathbf{x})\star {%
\hat{D}}_{\sigma \nu }^{\text{cl}}(\mathbf{x})=\frac{1}{2\pi \theta }{\hat{D}%
}_{\mu \nu }^{\text{cl}}(\mathbf{x}).  \label{ConstOnD}
\end{equation}%
Furthermore, we resolve the above noncommutative constraint by introducing
the noncommutative CP$^{N-1}$ field $n_{\mu }(\mathbf{x})$ with its complex
conjugate $\bar{n}_{\mu }(\mathbf{x})$, 
\begin{equation}
{\hat{D}}_{\mu \nu }^{\text{cl}}(\mathbf{x})=\frac{1}{2\pi \theta }n_{\mu }(%
\mathbf{x})\star \bar{n}_{\nu }(\mathbf{x}).  \label{DinN}
\end{equation}%
Recall that the noncommutative CP field has so far been introduced by hands
as a mere generalization of the ordinary (commutative) CP field\cite%
{LeePLB01}.

We then define the topological charge density by the formula 
\begin{equation}
J_{0}(\mathbf{x})=\frac{1}{2\pi \theta }\sum_{\mu }[\bar{n}_{\mu }(\mathbf{x}%
),n_{\mu }(\mathbf{x})]_{\star }.  \label{TopolChargDensi}
\end{equation}%
It follows from (\ref{DinN}) and (\ref{TopolChargDensi}) that the density
excitation $\Delta \rho ^{\text{cl}}(\mathbf{x})$ is essentially the
topological charge density,%
\begin{equation}
\Delta \rho ^{\text{cl}}(\mathbf{x})=-J_{0}(\mathbf{x}).  \label{ChargRelat}
\end{equation}%
It is a novel property of a noncommutative soliton that it induces an
excitation of the electron number density around it as dictated by this
formula. There are many different but equivalent ways of defining the
topological charge, but only this definition has such a special property.

An immediate consequence is that a topological soliton carries necessarily
the electron number $\Delta N_{\text{e}}^{\text{cl}}=-Q$, where $\Delta N_{%
\text{e}}^{\text{cl}}=\int \!d^{2}x\,\Delta \rho ^{\text{cl}}(\mathbf{x})$
and $Q=\int \!d^{2}x\,J_{0}(\mathbf{x})$. We should mention that this
property has been known\cite{Sondhi93B,Moon95B,Ezawa99L} since the first
proposal of skyrmions in QH systems. Nevertheless, there has been no
observation that the relation (\ref{ChargRelat}) holds rigorously with the
choice of the topological charge density (\ref{TopolChargDensi}).

Another remarkable result is that the skyrmion carrying $Q=1$ is constructed
as a W$_{\infty }$(N)-rotated state of a hole state. The corresponding CP$%
^{N-1}$ field is given by 
\begin{equation}
\mathfrak{n}_{\mu }=\sum_{n=0}^{\infty }\left[ u_{\mu }(n)|n\rangle \langle
n|+v_{\mu }(n)|n+1\rangle \langle n|\right] ,  \label{GenerSkyrm}
\end{equation}%
where $\mathfrak{n}_{\mu }$ is the Weyl-ordered operator whose symbol is $%
n_{\mu }(\mathbf{x})$, while $u_{\mu }(n)$ and $v_{\mu }(n)$ are infinitely
many parameters charactering the skyrmion. These parameters are fixed once
the Hamiltonian is given.

As an example we study the case where the Hamiltonian is given by the
hard-core four-fermion interaction\cite{LeeD01}. Determining those
parameters explicitly we obtain the skyrmion state as an eigenstate of the
Hamiltonian. The solution is found to possess a factorizable property, $%
S_{a}^{\text{cl}}(\mathbf{x})=\rho ^{\text{cl}}(\mathbf{x})\mathcal{S}_{a}(%
\mathbf{x})$, where $\mathcal{S}_{a}(\mathbf{x})$ is the solution in the
ordinary CP$^{N-1}$ model. We call such an isospin texture the factorizable
skyrmion. For the sake of completeness we analyze the skyrmion solution in
the noncommutative CP$^{N-1}$ model\cite{LeePLB01}. It is intriguing that
skyrmions found in the hard-core model\cite{LeeD01} and the noncommutative CP%
$^{N-1}$ model\cite{LeePLB01} are the same one though these two models are
very different. Finally we point out that a factorizable skyrmion cannot be
a physical excitation in the Hamiltonian system consisting of the Coulomb
and Zeeman interactions.

This paper is organized as follows. In section II we recapture the basic
moments of noncommutative geometry. In section III we review the properties
of the density operator for electrons in the lowest Landau level. In section
IV we discuss the topological charge in the noncommutative plane. In section
V we consider the four-fermion repulsive interaction governing the dynamics
of electrons in the lowest Landau level. In section VI we analyze the
skyrmion as a W$_{\infty }$(N)-rotated state of a hole state. In Section VII
we construct the microscopic skyrmion state in the hard-core model. In
section VIII we analyze the noncommutative CP$^{N-1}$ model. In section IX
we derive the effective theory for the classical density $\hat{D}_{\mu \nu
}^{\text{cl}}(\mathbf{x})$ and study a generic structure of classical
equations of motion. We also make the derivative expansion of the energy and
derive the noncommutative CP$^{N-1}$ model as the lowest order term. Section
X is devoted to discussions. In particular we briefly summarize the property
of skyrmions in the realistic Coulomb interaction model.

\section{Noncommutative Geometry}

The essence of noncommutative geometry becomes clearer when formulated in
algebraic terms\cite{Harvey}. Commutativity of a plane implies the algebra
of smooth functions over the plane, with the algebraic operation to be the
ordinary multiplication. The plane with the coordinate $\mathbf{x}=(x,y)$ is
said to be noncommutative if the algebraic operation is defined by 
\begin{equation}
f(\mathbf{x})\star h(\mathbf{x})=e^{-\frac{i}{2}\theta \nabla _{\mathbf{x}%
}\wedge \nabla _{\mathbf{y}}}f(\mathbf{x})h(\mathbf{y})|_{\mathbf{y}=\mathbf{%
x}},  \label{StarProdu}
\end{equation}%
where $\theta $ is the parameter of noncommutativity. The derivative is%
\begin{equation}
\partial _{i}f(\mathbf{x})=-\frac{i}{\theta }\epsilon _{ij}\left[ x_{j},f(%
\mathbf{x})\right] _{\star },  \label{DerivNCs}
\end{equation}%
where $\left[ f,g\right] _{\star }\equiv f\star g-g\star f$ is the Moyal
bracket. The $\star $-product (\ref{StarProdu}) is known to be the only
possible deformation of the ordinary product provided the associativity is
required.

From (\ref{StarProdu}) we get

\begin{equation}
x\star y-y\star x=-i\theta ,  \label{StarProduXY}
\end{equation}%
implying that the coordinates of a plane are noncommutative with respect to
the algebraic multiplication law.

We introduce operators $X$ and $Y$ forming the oscillator algebra,%
\begin{equation}
XY-YX=-i\theta .  \label{AlgebNC}
\end{equation}%
The mapping $(x,y)\mapsto (X,Y)$ generates that of a function $f(\mathbf{x})$
into the corresponding operator $O[f]$, which is the Weyl-ordered operator%
\cite{Weyl},%
\begin{equation}
O[f]=\frac{1}{2\pi }\int \!d^{2}k\,e^{i\mathbf{kX}}f(\mathbf{k}),
\label{WeylOpera}
\end{equation}%
where $f(\mathbf{k})$ is the Fourier transformation of $f(\mathbf{x})$, 
\begin{equation}
f(\mathbf{k})=\frac{1}{2\pi }\int \!d^{2}x\,e^{-i\mathbf{kx}}f(\mathbf{x}).
\end{equation}%
The function $f(\mathbf{x})$ is referred to as the symbol of $O[f]$.

The inverse of (\ref{WeylOpera}) is given by 
\begin{equation}
f(\mathbf{k})=\theta \text{Tr}\left( e^{-i\mathbf{kX}}O[f]\right) ,
\label{WeylOperaInver}
\end{equation}%
from which we find%
\begin{equation}
\int \!d^{2}x\,f(\mathbf{x})=2\pi \theta \text{Tr}\left( O[f]\right) .
\label{TraceInteg}
\end{equation}%
It follows from (\ref{WeylOpera}) that

\begin{equation}
O[f]O[g]=O[f\star g],
\end{equation}%
which generalizes the correspondence between (\ref{StarProduXY}) and (\ref%
{AlgebNC}).

The creation and annihilation operators are constructed from the algebra (%
\ref{AlgebNC}),

\begin{equation}
b=\frac{X-iY}{\sqrt{2\theta }},\quad \quad b^{\dag }=\frac{X+iY}{\sqrt{%
2\theta }},
\end{equation}%
generating the holomorphic basis,%
\begin{equation}
|n\rangle =\frac{(b^{\dag })^{n}}{\sqrt{n!}}|0\rangle ,\quad \quad
b|0\rangle =0,  \label{FockState}
\end{equation}%
in the Fock space. The Weyl-ordered operator $O[f]$ is presented as%
\begin{equation}
O[f]=\sum_{mn}O_{mn}[f]|m\rangle \langle n|,
\end{equation}%
where the matrix $O_{mn}[f]$ is given by

\begin{equation}
O_{mn}[f]=\langle m|O[f]|n\rangle .  \label{fmn}
\end{equation}%
The inversion formula (\ref{WeylOperaInver}) reads%
\begin{equation}
f(\mathbf{k})=\theta \sum_{mn}\langle n|e^{-i\mathbf{kX}}|m\rangle O_{mn}[f].
\label{RepreWeylD}
\end{equation}%
We have the chain of one-to-one mappings%
\begin{equation}
f\quad \leftrightarrow \quad O[f]\quad \leftrightarrow \quad O_{mn}[f],
\end{equation}%
accompanied by the ones among the multiplication laws%
\begin{equation}
f\star h\quad \leftrightarrow \quad O[f]O[h]\quad \leftrightarrow \quad
\sum_{j}O_{mj}[f]O_{jn}[h].
\end{equation}%
In the subsequent sections we will reproduce the above constructions in
terms of expectation values of field operators in the QH system.

\section{Electrons in the Lowest Landau Level}

The system of electrons in the lowest Landau level provides us with the
simplest and concrete example where the principle of noncommutativity
acquires a natural realization. In this section we recollect the basic
moments and introduce the bare operators which play the key role in the
consequent development. We assume the electron to carry the U(N) isospin
index. We have $N=2$ for the monolayer QH system with the spin degree of
freedom, and $N=4$ for the bilayer QH system with the spin and layer degrees
of freedom\cite{BookEzawa}.

\subsection{Quantum Mechanics}

A planar electron performs cyclotron motion in homogeneous perpendicular
magnetic field $(A_{i}^{\text{ext}}=\frac{1}{2}B_{\perp }\epsilon
_{ij}x_{j}) $. The electron coordinate $\mathbf{x}=(x,y)$ is decomposed into
the guiding center $\mathbf{X}=(X,Y)$ and the relative coordinate $\mathbf{R}%
=(R_{x},R_{y})$, $\mathbf{x}=\mathbf{X}+\mathbf{R}$, where $R_{i}=-\epsilon
_{ij}P_{j}/eB_{\perp }$ with $\mathbf{P}=(P_{x},P_{y})$ the covariant
momentum.

In the first quantized picture the operators $\mathbf{X}$ and $\mathbf{R}$
are 
\begin{subequations}
\label{xXR}
\begin{align}
X_{i}& =\frac{1}{2}x_{i}-i\ell _{B}^{2}\epsilon _{ij}\partial _{j}, \\
R_{i}& =\frac{1}{2}x_{i}+i\ell _{B}^{2}\epsilon _{ij}\partial _{j}.
\end{align}%
The canonical commutation relation implies 
\end{subequations}
\begin{subequations}
\begin{align}
\lbrack X_{i},X_{j}]& =-i\ell _{B}^{2}\epsilon _{ij},  \label{CommuXX} \\
\lbrack P_{i},P_{j}]& =i{\frac{\hbar ^{2}}{\ell _{B}^{2}}}\epsilon _{ij},
\label{CommuRR} \\
\lbrack X_{i},P_{j}]& =0,
\end{align}%
where $\ell _{B}$ is the magnetic length defined by $\ell _{B}^{2}=\hbar
/eB_{\perp }$.

The kinetic Hamiltonian with the electron mass $M$, 
\end{subequations}
\begin{equation}
H_{\text{K}}=\frac{\mathbf{P}^{2}}{2M}=\frac{1}{2M}%
(P_{x}-iP_{y})(P_{x}+iP_{y})+\frac{1}{2}\hbar \omega _{\text{c}},
\label{KinetHamil}
\end{equation}%
creates the equidistant Landau levels with gap energy $\hbar \omega _{\text{c%
}}=\hbar eB_{\perp }{/}M$.

Due to the noncommutative relation (\ref{CommuXX}) an electron cannot be
localized to a point and occupies an area $2\pi \ell _{B}^{2}$ in each
Landau level. It is highly degenerate. The degree of degeneracy is given by
the maximal affordable density of identical electrons given by $\rho _{\Phi
}=(2\pi \ell _{B}^{2})^{-1}$. It is equal to the magnetic flux density, $%
\rho _{\Phi }=B_{\perp }/\Phi _{\text{D}}$ with $\Phi _{\text{D}}=2\pi \hbar
/e$ the Dirac flux quantum. The maximal possible density is given by $N\rho
_{\Phi }$ for electrons carrying the isospin index $\mu =1,\cdots ,N$.

The algebra (\ref{CommuXX}) acts independently within each level. Provided
the magnetic field is strong enough, the gap energy $\hbar \omega _{\text{c}%
} $ becomes large compared with other characteristic energies associated
with thermal fluctuations or electrostatic interactions. Then, excitations
across Landau levels are practically suppressed, and electrons turn out to
be confined to the lowest Landau level. Namely, the degree of freedom
associated with the algebra (\ref{CommuRR}) is frozen, and the kinematic
Hamiltonian is quenched. This is the lowest-Landau-level (LLL) projection%
\cite{Girvin84B,Girvin85L}.

Consequently, the kinematics of the system becomes governed solely by the
algebra (\ref{CommuXX}), which is identical to (\ref{AlgebNC}). The
coordinates $x$ and $y$ standing in (\ref{StarProduXY}) appear in (\ref{xXR}%
) as coordinates in the space of representation of (\ref{CommuXX}). The
parameter of noncommutativity is $\theta =\ell _{B}^{2}$ and is carried
through all the subsequent account.

The wave functions of electrons in the lowest Landau level are given by%
\begin{equation}
\langle \mathbf{x}|n\rangle =\frac{z^{n}}{\sqrt{2^{1+n}\pi \theta n!}}%
e^{-|z|^{2}/4}  \label{WaveLLL}
\end{equation}%
with $z=(x+iy)/\sqrt{\theta }$, where $n$ labels the degenerate states with
respect to the orbital momentum eigenvalue. We call the state $|n\rangle $
the Landau site. We denote the number of Landau sites by $N_{\Phi }$, which
is equal to the number of flux quanta passing through the system. The set (%
\ref{WaveLLL}) agrees with the $\mathbf{x}$-representation of the states (%
\ref{FockState}). Though $\langle \mathbf{x}|n\rangle $ are orthonormal one
to another, their set is not complete, leading to the nonlocalizability of
an electron to a point when it is confined to the lowest Landau level.

\subsection{Second Quantization and Bare Densities}

In constructing the second quantized picture we introduce the fermion field
operator%
\begin{equation}
\psi _{\mu }(\mathbf{x})=\sum_{n=0}^{\infty }\langle \mathbf{x}|n\rangle
c_{\mu }(n),  \label{ElectInLLL}
\end{equation}%
where $\mu =1,\ldots ,N$ is the isospin index associated with the algebra
U(N). The fermion operators satisfy the standard anticommutation relations, 
\begin{equation}
\{c_{\mu }(m),c_{\nu }^{\dag }(n)\}=\delta _{\mu \nu }\delta _{mn}.
\label{AntiCCR}
\end{equation}%
The operator $c_{\mu }^{\dagger }(n)$ creates an electron with the isospin $%
\mu $ in the Landau site $n$.

The physical variables are the number density $\rho (\mathbf{x})=\psi ^{\dag
}(\mathbf{x})\psi (\mathbf{x})$ and the isospin density $S_{a}(\mathbf{x})=%
\frac{1}{2}\psi ^{\dag }(\mathbf{x})\lambda _{a}\psi (\mathbf{x})$ with the
Gell-Mann matrix $\lambda _{a}$. It is convenient to introduce the density
operator,%
\begin{equation}
D_{\mu \nu }(\mathbf{x})\equiv \psi _{\nu }^{\dag }(\mathbf{x})\psi _{\mu }(%
\mathbf{x}),  \label{DensiDz}
\end{equation}%
comprising the number and isospin densities as%
\begin{equation}
D_{\mu \nu }(\mathbf{x})=\frac{1}{N}\delta _{\mu \nu }\rho (\mathbf{x}%
)+(\lambda _{a})_{\mu \nu }S_{a}(\mathbf{x}).
\end{equation}%
Substituting (\ref{ElectInLLL}) into (\ref{DensiDz}), we obtain%
\begin{equation}
D_{\mu \nu }(\mathbf{x})=\sum_{mn}\langle n|\mathbf{x}\rangle \langle 
\mathbf{x}|m\rangle D_{\mu \nu }(m,n),
\end{equation}%
where%
\begin{equation}
D_{\mu \nu }(m,n)\equiv c_{\nu }^{\dag }(n)c_{\mu }(m).  \label{Dmn}
\end{equation}%
Using the relation%
\begin{equation}
\int d^{2}x\langle n|\mathbf{x}\rangle \langle \mathbf{x}|m\rangle e^{-i%
\mathbf{kx}}=e^{-\frac{1}{4}\theta k^{2}}\langle n|e^{-i\mathbf{kX}%
}|m\rangle ,
\end{equation}%
we get%
\begin{equation}
D_{\mu \nu }(\mathbf{x})=\frac{1}{2\pi }\int d^{2}k\,e^{-\frac{1}{4}\theta
k^{2}}{\hat{D}}_{\mu \nu }(\mathbf{k})e^{i\mathbf{kx}},  \label{DensiDy}
\end{equation}%
where%
\begin{equation}
{\hat{D}}_{\mu \nu }(\mathbf{k})=\frac{1}{2\pi }\sum_{mn}\langle n|e^{-i%
\mathbf{kX}}|m\rangle D_{\mu \nu }(m,n).  \label{DensiD}
\end{equation}%
Its Fourier transform reads%
\begin{equation}
{\hat{D}}_{\mu \nu }(\mathbf{x})=\frac{1}{2\pi }\int d^{2}k\,{\hat{D}}_{\mu
\nu }(\mathbf{k})e^{i\mathbf{kx}},  \label{DensiDx}
\end{equation}%
which is related to ${D}_{\mu \nu }(\mathbf{x})$ as%
\begin{equation}
D_{\mu \nu }(\mathbf{x})=\frac{1}{\pi \theta }\int d^{2}ye^{-|\mathbf{x}-%
\mathbf{y}|^{2}/\theta }{\hat{D}}_{\mu \nu }(\mathbf{y}).  \label{DhatD}
\end{equation}%
We call ${\hat{D}}_{\mu \nu }$ the bare density. This relation implies that
the physical density $D_{\mu \nu }(\mathbf{x})$ is not localizable to a
point.

Integrating both sides of (\ref{DhatD}) we get%
\begin{equation}
\int d^{2}xD_{\mu \nu }(\mathbf{x})=\int d^{2}x{\hat{D}}_{\mu \nu }(\mathbf{x%
}),
\end{equation}%
which indicates that (\ref{DhatD}) is just a smearing of ${\hat{D}}_{\mu \nu
}(\mathbf{x})$ over some area of order of $\theta $. The operator (\ref%
{DensiDx}) may be regarded to describe some kind of cores located inside the
physical objects associated with $D_{\mu \nu }(\mathbf{x})$.

\subsection{W$_{\infty }$(N) Algebra}

The algebraic relation 
\begin{align}
\lbrack D_{\mu \nu }(m,n),D_{\sigma \tau }(s,t)]=& \delta _{\mu \tau }\delta
_{mt}D_{\sigma \nu }(s,n)  \notag \\
& -\delta _{\sigma \nu }\delta _{sn}D_{\mu \tau }(m,t)  \label{WalgebD}
\end{align}%
hold between the density matrix operator, as is easily derived from the
anticommutation relation (\ref{AntiCCR}). We combine this with the
magnetic-translation group property,%
\begin{equation}
e^{i\mathbf{kX}}e^{i\mathbf{k}^{\prime }\mathbf{X}}=e^{i(\mathbf{k}+\mathbf{k%
}^{\prime })\mathbf{X}}\exp \left[ \frac{i}{2}\theta \mathbf{k}\!\wedge \!%
\mathbf{k}^{\prime }\right] ,  \label{MagneTrans}
\end{equation}%
which summarizes the noncommutativity of the plane. In this way we obtain%
\begin{align}
& 2\pi \lbrack {\hat{D}}_{\mu \nu }(\mathbf{k}),{\hat{D}}_{\sigma \tau }(%
\mathbf{k}^{\prime })]  \notag \\
& =\delta _{\mu \tau }e^{+\frac{i}{2}\theta \mathbf{k}\wedge \mathbf{k}%
^{\prime }}{\hat{D}}_{\sigma \nu }(\mathbf{k}+\mathbf{k}^{\prime })-\delta
_{\sigma \nu }e^{-\frac{i}{2}\theta \mathbf{k}\wedge \mathbf{k}^{\prime }}{%
\hat{D}}_{\mu \tau }(\mathbf{k}+\mathbf{k}^{\prime })  \label{WalgebP}
\end{align}%
in terms of the bare operator (\ref{DensiD}). We may rewrite it as 
\begin{subequations}
\begin{align}
\lbrack \hat{\rho}(\mathbf{k}),\hat{\rho}(\mathbf{k}^{\prime })]=& {\frac{i}{%
\pi }}\hat{\rho}(\mathbf{k}+\mathbf{k}^{\prime })\sin \left( \theta {\frac{%
\mathbf{k}\!\wedge \!\mathbf{k}^{\prime }}{2}}\right) ,  \label{SUCommu:A} \\
\lbrack \hat{S}_{a}(\mathbf{k}),\hat{\rho}(\mathbf{k}^{\prime })]=& {\frac{i%
}{\pi }}\hat{S}_{a}(\mathbf{k}+\mathbf{k}^{\prime })\sin \left( \theta {%
\frac{\mathbf{k}\!\wedge \!\mathbf{k}^{\prime }}{2}}\right) ,
\label{SUCommu:B} \\
\lbrack \hat{S}_{a}(\mathbf{k}),\hat{S}_{b}(\mathbf{k}^{\prime })]=& {\frac{i%
}{2\pi }}f_{abc}\hat{S}_{c}(\mathbf{k}+\mathbf{k}^{\prime })\cos \left(
\theta {\frac{\mathbf{k}\!\wedge \!\mathbf{k}^{\prime }}{2}}\right)  \notag
\\
& +{\frac{i}{2\pi }}d_{abc}\hat{S}_{c}(\mathbf{k}+\mathbf{k}^{\prime })\sin
\left( \theta {\frac{\mathbf{k}\!\wedge \!\mathbf{k}^{\prime }}{2}}\right) 
\notag \\
& +{\frac{i}{2\pi N}}\delta _{ab}\hat{\rho}(\mathbf{k}+\mathbf{k}^{\prime
})\sin \left( \theta {\frac{\mathbf{k}\!\wedge \!\mathbf{k}^{\prime }}{2}}%
\right) ,  \label{SUCommu:C}
\end{align}%
where the summation over the repeated isospin index is understood. We have
named this the W$_{\infty }$(N) algebra\cite{EzawaX03B} since it is the
SU(N) extension of W$_{\infty }$. Its physical implication is an intrinsic
entanglement between the electron density and the isospin density. The
entanglement is removed in the commutative limit, $\theta \rightarrow 0$,
where it is reduced to the U(1)$\otimes $SU(N) algebra.

\section{Noncommutative Kinematics}

In this section we study how the noncommutativity presented in microscopic
theory generates a constraint on classical objects. We first define the
class of Fock states we work with and then derive a noncommutative relation
satisfied by the expectation value of the bare density. We subsequently
introduce the noncommutative CP$^{N-1}$ field by resolving the
noncommutative constraint, and discuss topological aspects with the use of
the noncommutative CP$^{N-1}$ field.

\subsection{Fock States}

We consider the class of Fock states which can be written as 
\end{subequations}
\begin{equation}
|\mathfrak{S}\rangle =e^{iW}|\mathfrak{S}_{0}\rangle ,  \label{SkyrmFormuC}
\end{equation}%
where $W$ is an arbitrary element of the algebra $W_{\infty }(N)$ which
represents a general linear combination of the operators (\ref{Dmn}). The
state $|\mathfrak{S}_{0}\rangle $ is assumed to be of the form

\begin{equation}
|\mathfrak{S}_{0}\rangle =\prod_{\mu ,n}\left[ c_{\mu }^{\dag }(n)\right]
^{\nu _{\mu }(n)}|0\rangle ,  \label{SkyrmFormuCx}
\end{equation}%
where $\nu _{\mu }(n)$ takes the value either $0$ or $1$ specifying whether
the isospin state $\mu $ at the Landau site $n$ is occupied or not,
respectively.

The amount of electrons at the Landau site $n$ is given by%
\begin{equation}
\nu (n)=\sum_{\mu =1}^{N}\nu _{\mu }(n),
\end{equation}%
and may take a value from $0$ up to $N$. The filling factor is defined by%
\begin{equation}
\nu =\frac{1}{N_{\Phi }}\sum_{m=0}^{N_{\Phi }-1}\nu (m),
\end{equation}%
where the thermodynamical limit $N_{\Phi }\rightarrow \infty $ is implied.
The electron density is homogeneous in the ground state, $\nu (n)=$constant.

The electron number of the W$_{\infty }$(N)-rotated state (\ref{SkyrmFormuC}%
) is easily calculable,%
\begin{equation}
\langle \mathfrak{S}|N_{\text{e}}|\mathfrak{S}\rangle =\langle \mathfrak{S}%
_{0}|e^{-iW}N_{\text{e}}e^{+iW}|\mathfrak{S}_{0}\rangle =\langle \mathfrak{S}%
_{0}|N_{\text{e}}|\mathfrak{S}_{0}\rangle ,
\end{equation}%
since the total electron number 
\begin{equation}
N_{\text{e}}=\sum_{n}\sum_{\mu }c_{\mu }^{\dagger }(n)c_{\mu }(n)
\end{equation}%
is a Casimir operator. The electron number of the state $|\mathfrak{S}%
\rangle $ is the same as that of the state $|\mathfrak{S}_{0}\rangle $. We
set%
\begin{equation}
\Delta N_{\text{e}}^{\text{cl}}=\langle \mathfrak{S}|N_{\text{e}}|\mathfrak{S%
}\rangle -\langle \text{g}|N_{\text{e}}|\text{g}\rangle =\int
\!d^{2}x\,\Delta \hat{\rho}^{\text{cl}}(\mathbf{x}),  \label{ElectNumbeA}
\end{equation}%
where $\langle $g$|N_{\text{e}}|$g$\rangle $ is the total electron number in
the ground state $|$g$\rangle $, and $\Delta N_{\text{e}}^{\text{cl}}$ is
the number of extra electrons carried by the excitation described by the
state $|\mathfrak{S}\rangle $. It is an integer since $\langle \mathfrak{S}%
_{0}|N_{e}|\mathfrak{S}_{0}\rangle $ is an integer as well as $\langle $g$%
|N_{\text{e}}|$g$\rangle $.

The class of states of the form (\ref{SkyrmFormuC}) together with (\ref%
{SkyrmFormuCx}) does not embrace the whole Fock space. Nevertheless, it is
general enough to cover all physically relevant cases in our analysis where
the filling factor $\nu $ is taken to be an integer. Indeed, as far as we
know, perturbative excitations are isospin waves and nonperturbative
excitations are skyrmions in QH systems. The corresponding states belong
surely to this category.

\subsection{Noncommutative Constraints}

The classical field is the expectation value of a second quantized operator
such as 
\begin{equation}
\hat{D}_{\mu \nu }^{\text{cl}}(\mathbf{x})\equiv \langle \mathfrak{S}|{\hat{D%
}}_{\mu \nu }(\mathbf{x})|\mathfrak{S}\rangle .
\end{equation}%
From (\ref{DensiD}) we have 
\begin{equation}
\hat{D}_{\mu \nu }^{\text{cl}}(\mathbf{k})={\frac{1}{2\pi }}\sum_{mn}\langle
m|e^{-i\mathbf{kX}}|n\rangle D_{\mu \nu }^{\text{cl}}(m,n),
\label{ProjeDensiClass}
\end{equation}%
where 
\begin{equation}
D_{\mu \nu }^{\text{cl}}(m,n)=\langle \mathfrak{S}|c_{\nu }^{\dag }(n)c_{\mu
}(m)|\mathfrak{S}\rangle .
\end{equation}%
It is identical to (\ref{RepreWeylD}) when we set $f(\mathbf{k})=\hat{D}%
_{\mu \nu }^{\text{cl}}(\mathbf{k})$ and $O_{mn}[f]=D_{\mu \nu }^{\text{cl}%
}(m,n)/2\pi \theta $. Hence, from the matrix element $D_{\mu \nu }^{\text{cl}%
}(m,n)$ we may construct a Weyl-ordered operator whose symbol is $\hat{D}%
_{\mu \nu }^{\text{cl}}(\mathbf{x})$.

As we derive in Appendix \ref{AppenA} the classical density satisfies the
relation%
\begin{equation}
\sum_{\sigma ,s}D_{\mu \sigma }^{\text{cl}}(m,s)D_{\sigma \nu }^{\text{cl}%
}(s,n)=D_{\mu \nu }^{\text{cl}}(m,n).  \label{DDD}
\end{equation}%
In terms of symbols it reads%
\begin{equation}
\sum_{\sigma =1}^{N}{\hat{D}}_{\mu \sigma }^{\text{cl}}(\mathbf{x})\star {%
\hat{D}}_{\sigma \nu }^{\text{cl}}(\mathbf{x})=\frac{1}{2\pi \theta }{\hat{D}%
}_{\mu \nu }^{\text{cl}}(\mathbf{x}).  \label{ConstNCa}
\end{equation}%
The noncommutativity encoded in (\ref{CommuXX}) has become perceptible at
the classical level in terms of the bare quantity $\hat{D}_{\mu \nu }^{\text{%
cl}}(\mathbf{x})$. This is the noncommutative constraint on the bare density.

We may rewrite (\ref{ConstNCa}) as the constraints on the bare densities ${%
\hat{\rho}}^{\text{cl}}(\mathbf{x})$ and $\hat{S}_{a}^{\text{cl}}(\mathbf{x}%
) $ given by 
\begin{equation}
{\hat{D}}_{\mu \nu }^{\text{cl}}(\mathbf{x})=\frac{1}{N}\delta _{\mu \nu }{%
\hat{\rho}}^{\text{cl}}(\mathbf{x})+(\lambda _{a})_{\mu \nu }\hat{S}_{a}^{%
\text{cl}}(\mathbf{x}),  \label{DrhoSclass}
\end{equation}%
where ${\hat{\rho}}^{\text{cl}}(\mathbf{x})\equiv \langle \mathfrak{S}|{\hat{%
\rho}}(\mathbf{x})|\mathfrak{S}\rangle $ and $\hat{S}_{a}^{\text{cl}}(%
\mathbf{x})\equiv \langle \mathfrak{S}|\hat{S}_{a}(\mathbf{x})|\mathfrak{S}%
\rangle $. Substituting this into (\ref{ConstNCa}) and using the properties
of Gell-Mann matrices we come to 
\begin{subequations}
\label{ConstNCb}
\begin{align}
& \hat{S}_{a}^{\text{cl}}(\mathbf{x})\star \hat{S}_{a}^{\text{cl}}(\mathbf{x}%
)=\frac{1}{4\pi \theta }{\hat{\rho}}^{\text{cl}}(\mathbf{x})-\frac{1}{2N}{%
\hat{\rho}}^{\text{cl}}(\mathbf{x})\star {\hat{\rho}}^{\text{cl}}(\mathbf{x}%
), \\
& \frac{1}{2}(if_{abc}+d_{abc})\hat{S}_{b}^{\text{cl}}(\mathbf{x})\star \hat{%
S}_{c}^{\text{cl}}(\mathbf{x})  \notag \\
& =\frac{1}{4\pi \theta }\hat{S}_{a}^{\text{cl}}(\mathbf{x})-\frac{1}{2N}{%
\hat{\rho}}^{\text{cl}}(\mathbf{x})\star \hat{S}_{a}^{\text{cl}}(\mathbf{x})-%
\frac{1}{2N}\hat{S}_{a}^{\text{cl}}(\mathbf{x})\star {\hat{\rho}}^{\text{cl}%
}(\mathbf{x}).
\end{align}%
These two relations are the manifestation of the microscopic
noncommutativity at the level of the classical fields.

\subsection{Noncommutative CP$^{N-1}$ Field}

Due to the involution of the $\star $-product in (\ref{ConstNCb}), the
structure of the fields ${\hat{\rho}}^{\text{cl}}(\mathbf{x})$ and $\hat{S}%
_{a}^{\text{cl}}(\mathbf{x})$ is very complicated, and a comprehensive
analysis of these equations is quite far from being trivial. We investigate
relatively simple cases of integer filling factors. In the physics of QH
systems the integer value of $\nu $ is related with the integer quantum Hall
effect, which is much simpler in structure than the fractional one realized
at a certain fractional value of $\nu $. In this case we are able to present
a systematic way of resolving the noncommutative relations (\ref{ConstNCb})
or (\ref{ConstNCa}).

We introduce the $\nu $ amount $(r=1,\ldots ,\nu )$ of fields $n^{r}(\mathbf{%
x})$ each representing a column with $N\ $complex components. They are
required to be orthonormal 
\end{subequations}
\begin{equation}
\bar{n}_{\mu }^{r}(\mathbf{x})\star n_{\mu }^{s}(\mathbf{x})=\delta ^{rs},
\label{ConstCP}
\end{equation}%
where $\bar{n}$ denotes the complex conjugate of $n$. Here and hereafter the
summation over the repeated isospin index $\mu $ is understood. Then it is
trivial to see that the relation (\ref{ConstNCa}) is resolved by%
\begin{equation}
{\hat{D}}_{\mu \nu }^{\text{cl}}(\mathbf{x})=\frac{1}{2\pi \theta }%
\sum_{s=1}^{\nu }n_{\mu }^{s}(\mathbf{x})\star \bar{n}_{\nu }^{s}(\mathbf{x}%
),  \label{DinCP}
\end{equation}%
or 
\begin{subequations}
\label{RhoSinN}
\begin{align}
{\hat{\rho}}^{\text{cl}}(\mathbf{x})& =\frac{1}{2\pi \theta }\sum_{s}n_{\mu
}^{s}(\mathbf{x})\star \bar{n}_{\mu }^{s}(\mathbf{x}),  \label{RhoInN} \\
{\hat{S}}_{a}^{\text{cl}}(\mathbf{x})& =\frac{1}{4\pi \theta }%
\sum_{s}(\lambda _{a})_{\mu \nu }n_{\nu }^{s}(\mathbf{x})\star \bar{n}_{\mu
}^{s}(\mathbf{x}).  \label{SinN}
\end{align}%
Changing the order of multiplicatives in (\ref{RhoInN}) and using (\ref%
{ConstCP}) we obtain 
\end{subequations}
\begin{equation}
{\hat{\rho}}^{\text{cl}}(\mathbf{x})=\frac{\nu }{2\pi \theta }+\frac{1}{2\pi
\theta }\sum_{s}\left[ n_{\mu }^{s}(\mathbf{x}),\bar{n}_{\mu }^{s}(\mathbf{x}%
)\right] _{\star }.  \label{RhoInNx}
\end{equation}%
The first term represents the homogeneous part of the electron density. Note
that $\rho _{0}\equiv \nu \rho _{\Phi }=\nu /2\pi \theta $ is the electron
density in the ground state. Comparing this with (\ref{ElectNumbeA}) we find%
\begin{equation}
\Delta \hat{\rho}^{\text{cl}}(\mathbf{x})=\frac{1}{2\pi \theta }\sum_{s}%
\left[ n_{\mu }^{s}(\mathbf{x}),\bar{n}_{\mu }^{s}(\mathbf{x})\right]
_{\star },  \label{DensiExcit}
\end{equation}%
which is the density excitation from the ground state realized in the state (%
\ref{SkyrmFormuC}).

At the filling factor $\nu =1$, which we study mostly in what follows, we
have a single $N$-component column $n(\mathbf{x})$ satisfying

\begin{equation}
\bar{n}_{\mu }(\mathbf{x})\star n_{\mu }(\mathbf{x})=1,  \label{ConstCPone}
\end{equation}%
and the bare density reads%
\begin{equation}
{\hat{D}}_{\mu \nu }^{\text{cl}}(\mathbf{x})=\frac{1}{2\pi \theta }n_{\mu }(%
\mathbf{x})\star \bar{n}_{\nu }(\mathbf{x}),  \label{DinCPone}
\end{equation}%
or 
\begin{subequations}
\label{RhoSinNx}
\begin{align}
{\hat{\rho}}^{\text{cl}}(\mathbf{x})& =\frac{1}{2\pi \theta }n_{\mu }(%
\mathbf{x})\star \bar{n}_{\mu }(\mathbf{x}), \\
{\hat{S}}_{a}^{\text{cl}}(\mathbf{x})& =\frac{1}{4\pi \theta }(\lambda
_{a})_{\mu \nu }n_{\nu }(\mathbf{x})\star \bar{n}_{\mu }(\mathbf{x}).
\end{align}%
They are represented as 
\end{subequations}
\begin{equation}
\mathfrak{n}_{\mu }^{\dag }\mathfrak{n}_{\mu }=1,
\end{equation}%
and%
\begin{equation}
O[\hat{D}_{\mu \nu }^{\text{cl}}]=\frac{1}{2\pi \theta }\mathfrak{n}_{\mu }%
\mathfrak{n}_{\nu }^{\dag }  \label{DinCPw}
\end{equation}%
in terms of the Weyl-ordered operator $\mathfrak{n}_{\mu }$ whose symbol is
the CP$^{N-1}$ field $n_{\mu }(\mathbf{x})$, $\mathfrak{n}_{\mu }=O[n_{\mu
}] $.

We study a local noncommutative phase transformation of $n_{\mu }(\mathbf{x}%
) $. Let $U(\mathbf{x})$ be a complex valued function satisfying 
\begin{equation}
\bar{U}(\mathbf{x})\star U(\mathbf{x})=U(\mathbf{x})\star \bar{U}(\mathbf{x}%
)=1,
\end{equation}%
which is constructed as%
\begin{equation}
U=e_{\star }^{i\xi }\equiv 1+i\xi +\frac{i^{2}}{2!}\xi \star \xi +\cdots
\end{equation}%
with $\xi (\mathbf{x})$ being an arbitrary function. Then the local $U(1)$
transformation of $n_{\mu }(\mathbf{x})$ is given by 
\begin{subequations}
\begin{align}
n_{\mu }(\mathbf{x})& \rightarrow n_{\mu }^{\prime }(\mathbf{x})\equiv
n_{\mu }(\mathbf{x})\star U(\mathbf{x}), \\
\bar{n}_{\mu }(\mathbf{x})& \rightarrow \bar{n}_{\mu }^{\prime }(\mathbf{x}%
)\equiv \bar{U}(\mathbf{x})\star \bar{n}_{\mu }(\mathbf{x}),
\end{align}%
which are consistent as a matter of $\overline{(f\star h)}=\bar{h}\star \bar{%
f}$. The classical density (\ref{RhoSinNx}) as well as the noncommutative
normalization condition (\ref{ConstCPone}) are invariant under the
transformation.

In general, at the integer filling factor $\nu $, we introduce the
Grassmannian G$_{N,\nu }$ field\cite{macfarlain} as a set of the $\nu $
amount of the CP$^{N-1}$ fields, 
\end{subequations}
\begin{equation}
Z(\mathbf{x})=(n_{\mu }^{1},n_{\mu }^{2},\cdots ,n_{\mu }^{\nu }).
\end{equation}%
Under a local $U(\mathbf{x})$ transformation, 
\begin{equation}
Z(\mathbf{x})\rightarrow Z^{\prime }(\mathbf{x})\equiv Z(\mathbf{x})\star U(%
\mathbf{x}),
\end{equation}%
the classical density (\ref{RhoSinN}) as well as the noncommutative
normalization condition (\ref{ConstCP}) are invariant.

\subsection{Topological Charge}

There exist several equivalent ways of defining the topological charge
associated with the CP$^{N-1}$ field in the ordinary theory on the
commutative plane. One of them reads%
\begin{equation}
Q=\frac{1}{2\pi i}\epsilon _{kl}\sum_{s}\int \!d^{2}x\,(\partial _{k}\bar{n}%
_{\mu }^{s})(\partial _{l}n_{\mu }^{s}).  \label{TopolCPy}
\end{equation}%
We generalize it to the noncommutative theory as%
\begin{equation}
Q=\frac{1}{2\pi \theta }\sum_{s}\int \!d^{2}x\,[\bar{n}_{\mu }^{s}(\mathbf{x}%
),n_{\mu }^{s}(\mathbf{x})]_{\star },  \label{TopolChargNC}
\end{equation}%
since the Moyal bracket is expanded as%
\begin{equation}
\lbrack \bar{n}_{\mu }^{s}(\mathbf{x}),n_{\mu }^{s}(\mathbf{x})]_{\star
}=-i\theta \epsilon _{jk}\partial _{j}\bar{n}_{\mu }^{s}(\mathbf{x})\partial
_{k}n_{\mu }^{s}(\mathbf{x})+O(\theta ^{2}),
\end{equation}%
and the quantity (\ref{TopolChargNC}) is reduced to the topological charge (%
\ref{TopolCPy}) in the commutative limit ($\theta \rightarrow 0$).

We now show that the quantity (\ref{TopolChargNC}) satisfies all standard
requirements that the topological charge should meet. The topological charge
density is%
\begin{equation}
J_{0}(\mathbf{x})=\frac{1}{2\pi \theta }\sum_{s,\mu }[\bar{n}_{\mu }^{s}(%
\mathbf{x}),n_{\mu }^{s}(\mathbf{x})]_{\star }.  \label{TopolDensi}
\end{equation}%
First, due to an important property of the star product such that%
\begin{equation}
f(\mathbf{x})\star g(\mathbf{x})=f(\mathbf{x})g(\mathbf{x})+\partial
_{k}\Lambda _{k}(\mathbf{x}),  \label{StarPropeA}
\end{equation}%
the density is expressed as $J_{0}(\mathbf{x})=\partial _{k}\Lambda _{k}(%
\mathbf{x})$ with a certain function $\Lambda _{k}(\mathbf{x})$. Thus (\ref%
{TopolChargNC}) reads 
\begin{equation}
Q=\int \!d^{2}x\,J_{0}(\mathbf{x})=\oint \!dx_{k}\,\Lambda _{k}(\mathbf{x}),
\end{equation}%
where the contour integration is taken along a circle of an infinite radius.
At the spatial infinity, the normalization (\ref{ConstCPone}) turns into a
commutative condition (since the derivatives must vanish), and the
asymptotic behavior is given by $n_{\mu }(\mathbf{x})\sim r^{0}$, yielding $%
\Lambda _{k}\sim r^{-1}$. Thus, $Q\sim \oint \!d\vartheta $ with $\vartheta $
the azimuthal angle. The integration over the angle gives the number of
windings. This is the standard way how the number of windings appear.

We next examine the conservation law. We obtain%
\begin{align}
\partial _{0}J_{0}=& \frac{1}{2\pi \theta }\sum_{s,\mu }[\partial _{0}\bar{n}%
_{\mu }^{s}(\mathbf{x}),n_{\mu }^{s}(\mathbf{x})]_{\star }  \notag \\
& +\frac{1}{2\pi \theta }\sum_{s,\mu }[\bar{n}_{\mu }^{s}(\mathbf{x}%
),\partial _{0}n_{\mu }^{s}(\mathbf{x})]_{\star }=\partial _{k}J_{k},
\end{align}%
where the existence of $J_{k}$ is guaranteed by the property (\ref%
{StarPropeA}). Based on dimensional considerations, the quantity $J_{k}$
behaves as $J_{k}\sim r^{-2}$ asymptotically, implying the vanishing of the
corresponding surface integral, and the conservation of the topological
charge, $\partial _{0}Q=0$, holds.

As we have stated, there are several ways of defining the topological charge 
$Q$. Needless to say, they are all equivalent. Nevertheless the topological
charge densities are different, and in general they have no physical
meaning. Among them the topological charge density (\ref{TopolDensi}) has a
privileged role that it is essentially the electron density excitation
induced by the topological soliton,%
\begin{equation}
J_{0}(\mathbf{x})=-\Delta \hat{\rho}^{\text{cl}}(\mathbf{x}),
\label{TopolElectDensi}
\end{equation}%
as follows from the comparison of (\ref{TopolDensi}) with (\ref{DensiExcit}%
). Hence the topological charge is given by%
\begin{equation}
Q=-\int \!d^{2}x\,\Delta \hat{\rho}^{\text{cl}}(\mathbf{x})=-\Delta N_{\text{%
e}}^{\text{cl}},  \label{RelatChargNumbe}
\end{equation}%
which is an integer. Consequently the noncommutative soliton carries
necessarily the electric charge proportional to its topological charge. This
is a very peculiar phenomenon that occurs in the noncommutative theory.

\section{Hamiltonian}

We have so far been concerned with the kinematical properties of the
noncommutative electron system. Recall that the LLL projection quenches the
kinetic Hamiltonian (\ref{KinetHamil}) and brings about the noncommutativity
into the system. We now investigate interactions between electrons which
organize the system dynamically.

\subsection{Four-Fermion Interactions}

We take a four-fermion interaction term, 
\begin{equation}
H_{\text{V}}={\frac{1}{2}}\int \!d^{2}xd^{2}y\,V(\mathbf{x}-\mathbf{y}%
)\Delta \rho (\mathbf{x})\Delta \rho (\mathbf{y}),  \label{CouloHamil}
\end{equation}%
where $\Delta \rho (\mathbf{x})$ is the density excitation operator, 
\begin{equation}
\Delta \rho (\mathbf{x})=\rho (\mathbf{x})-\rho _{0},
\end{equation}%
and $\rho _{0}$ denotes the homogeneous density of charges providing the
electric neutrality in the ground state. In the real QH system $V(\mathbf{x}%
) $ is given by the Coulomb potential, but here we only assume that it
represents a repulsive interaction.

As we have already mentioned, we consider the system at an integer filling
factor $\nu $. Each Landau site can accommodate up to $N\ $electrons of
different isospins. The cases of $\nu =0$ and $\nu =N$ are trivial since the
isospin polarization is zero, where Landau sites are either all empty or all
occupied. Nontrivial filling factors are $\nu =1,\ldots ,N-1$. The fillings $%
\nu $ and $N-\nu $ are equivalent in the sense of particle-hole symmetry.

Substituting the density operators (\ref{ElectInLLL}) into (\ref{CouloHamil}%
) we obtain 
\begin{equation}
H=\sum_{mnij}V_{mnij}{D}_{\mu \mu }(n,m){D}_{\nu \nu }(j,i)-\nu \left( N_{%
\text{e}}+\Delta N_{\text{e}}\right) \epsilon _{\text{D}},
\label{HamilMonoNO}
\end{equation}%
where $N_{\text{e}}=\nu N_{\Phi }+\Delta N_{\text{e}}$, and%
\begin{equation}
{V}_{mnij}={\frac{1}{4\pi }}\int \!d^{2}k\,V(\mathbf{k})e^{-\theta \mathbf{k}%
^{2}/2}\langle m|e^{i\mathbf{kX}}|n\rangle \langle i|e^{-i\mathbf{kX}%
}|j\rangle ,  \label{Vmnij}
\end{equation}%
with $V(\mathbf{k})$ the Fourier transformation of the potential $V(\mathbf{x%
})$. We have used the notation 
\begin{align}
& \Delta N_{\text{e}}=\int \!d^{2}x\,\Delta \rho (\mathbf{x}),
\label{SpinNumbe} \\
& \epsilon _{\text{D}}=\sum_{j}{V}_{nnjj}=\frac{\rho _{0}}{2}\int
\!d^{2}x\,V(\mathbf{x}).  \label{EnergD}
\end{align}%
For later convenience we also define%
\begin{equation}
\epsilon _{\text{X}}=\sum_{j}{V}_{njjn}=\frac{\rho _{0}}{2}\theta \int
\!d^{2}k\,V(\mathbf{k})e^{-\theta \mathbf{k}^{2}/2}.  \label{EnergX}
\end{equation}%
Here, $\epsilon _{\text{D}}$ and $\epsilon _{\text{X}}$ are the direct and
exchange energy parameters, respectively. Note that $\sum_{j}{V}_{nnjj}$ and 
$\sum_{j}{V}_{njjn}$ are independent of $n$, and the summation is taken over 
$j$ with $n$ arbitrarily fixed.

\subsection{Spontaneous Symmetry Breaking}

Using the four-fermion interaction Hamiltonian (\ref{CouloHamil}) we
evaluate the energy of the state $|\mathfrak{S}\rangle $ in the class (\ref%
{SkyrmFormuC}). In Appendix A we derive the decomposition formula, 
\begin{equation}
E_{\text{V}}\equiv \langle \mathfrak{S}|H_{\text{V}}|\mathfrak{S}\rangle =E_{%
\text{D}}+E_{\text{X}},  \label{DecomFormu}
\end{equation}%
where 
\begin{subequations}
\label{EnergDX}
\begin{align}
E_{\text{D}}& =V_{mnij}D_{\mu \mu }^{\text{cl}}(n,m)D_{\nu \nu }^{\text{cl}%
}(j,i)-\nu (N_{\text{e}}^{\text{cl}}+\Delta N_{\text{e}}^{\text{cl}%
})\epsilon _{\text{D}},  \label{EnergDa} \\
E_{\text{X}}& =-V_{mnij}D_{\mu \nu }^{\text{cl}}(j,m)D_{\nu \mu }^{\text{cl}%
}(n,i)+N_{\text{e}}^{\text{cl}}\epsilon _{\text{X}},  \label{EnergXa}
\end{align}%
with $N_{\text{e}}^{\text{cl}}=\nu N_{\Phi }+\Delta N_{\text{e}}^{\text{cl}}$%
. Here $E_{\text{D}}$ and $E_{\text{X}}$ represent the direct and exchange
energies.

It is straightforward to represent them in the momentum space by using (\ref%
{ProjeDensiClass}) and (\ref{Vmnij}), 
\end{subequations}
\begin{subequations}
\label{DecomDXx}
\begin{align}
E_{\text{D}}=& \pi \int \!d^{2}k\,V(\mathbf{k})e^{-\frac{1}{2}\theta
k^{2}}\left\vert \Delta \hat{\rho}^{\text{cl}}(\mathbf{k})\right\vert ^{2},
\label{AppenSSBa} \\
E_{\text{X}}=& \pi \int \!d^{2}k\,\Delta V_{\text{X}}(\mathbf{k})\left[
\left\vert \hat{S}_{a}^{\text{cl}}(\mathbf{k})\right\vert ^{2}+\frac{1}{4}%
\left\vert \hat{\rho}^{\text{cl}}(\mathbf{k})\right\vert ^{2}\right] ,
\label{AppenSSBb}
\end{align}%
where $\Delta V_{\text{X}}(\mathbf{k})=V_{\text{X}}(0)-V_{\text{X}}(\mathbf{k%
})$ with 
\end{subequations}
\begin{equation}
V_{\text{X}}(\mathbf{k})\equiv \frac{\ell _{B}^{2}}{\pi }\int
\!d^{2}k\,e^{-i\theta \mathbf{k}\wedge \mathbf{k}^{\prime }}e^{-\frac{1}{2}%
\theta k^{\prime }{}^{2}}V(\mathbf{k}^{\prime }).
\end{equation}%
Note that both the energies are positive semidefinite since $V(\mathbf{k})>0$
and $\Delta V_{\text{X}}(\mathbf{k})>0$ for a repulsive interaction. Here, $%
\hat{\rho}^{\text{cl}}(\mathbf{k})=\langle \mathfrak{S}|\hat{\rho}(\mathbf{k}%
)|\mathfrak{S}\rangle $ and $\hat{S}_{a}^{\text{cl}}(\mathbf{k})=\langle 
\mathfrak{S}|\hat{S}_{a}(\mathbf{k})|\mathfrak{S}\rangle $. It is remarkable
that, though the Hamiltonian (\ref{CouloHamil}) involves no isospin
variables, the energy (\ref{DecomFormu}) of a state does. The direct energy $%
E_{\text{D}}$ is insensitive to isospin orientations, and it vanishes for
the homogeneous electron distribution since $\Delta \hat{\rho}^{\text{cl}}(%
\mathbf{k})=0$. The exchange energy $E_{\text{X}}$ depends on isospin
orientations. The isospin texture is homogeneous when the isospin is
completely polarized, where $\hat{S}_{a}^{\text{cl}}(\mathbf{k})\varpropto
\delta (\mathbf{k})$. Furthermore, $\hat{\rho}^{\text{cl}}(\mathbf{k}%
)\varpropto \delta (\mathbf{k})$ due to the homogeneous electron
distribution. For such a isospin orientation the exchange energy also
vanishes since $\Delta V_{\text{X}}(\mathbf{k})=V_{\text{X}}(0)-V_{\text{X}}(%
\mathbf{k})=0$ in (\ref{AppenSSBb}). On the other hand, $E_{\text{X}}>0$ if
the isospin is not polarized completely since $\hat{S}_{a}^{\text{cl}}(%
\mathbf{k})$ contains nonzero momentum components. Consequently the
isospin-polarized state has the lowest energy, which is zero. Namely, all
isospins are spontaneously polarized along an arbitrarily chosen direction.
It implies a spontaneous symmetry breaking due to the exchange interaction
implicit in the Hamiltonian (\ref{CouloHamil}).

\subsection{Ground State and Goldstone Modes}

The ground state is a spontaneously symmetry broken one, where the electron
density is homogenous, ${\rho }^{\text{cl}}(\mathbf{x})=\rho _{0}\equiv \nu
/2\pi \theta $, and the isospin field is additionally a constant, $\hat{S}%
_{a}^{\text{cl}}(\mathbf{x})=\mathcal{S}_{a}^{0}/2\pi \theta $. Substituting
these into the noncommutative constraint (\ref{ConstNCb}) we find 
\begin{subequations}
\begin{align}
\mathcal{S}_{a}^{0}\mathcal{S}_{a}^{0}& =\frac{\nu (N-\nu )}{2N}, \\
d_{abc}\mathcal{S}_{b}^{0}\mathcal{S}_{c}^{0}& =\frac{N-2\nu }{N}\mathcal{S}%
_{a}^{0}.
\end{align}%
At the filling factor $\nu $ the isospin is normalized on the ground state
in this way. The asymptotic behavior of the topological soliton is
determined by these equations.

We study small perturbative fluctuations of the isospin on the ground state.
We may keep the electron density unperturbed, ${\rho }^{\text{cl}}(\mathbf{x}%
)=\rho _{0}$, to minimize the direct energy (\ref{AppenSSBa}). Setting 
\end{subequations}
\begin{equation}
\hat{S}_{a}^{\text{cl}}(\mathbf{x})=\mathcal{S}_{a}(\mathbf{x})/2\pi \theta ,
\end{equation}%
and substituting these into (\ref{ConstNCb}), we find 
\begin{subequations}
\begin{align}
\mathcal{S}_{a}(\mathbf{x})\star \mathcal{S}_{a}(\mathbf{x})& =\frac{\nu
(N-\nu )}{2N}, \\
(if_{abc}+d_{abc})\mathcal{S}_{b}(\mathbf{x})\star \mathcal{S}_{c}(\mathbf{x}%
)& =\frac{N-2\nu }{N}\mathcal{S}_{a}(\mathbf{x}).
\end{align}%
The Goldstone modes, which are small fluctuations of the isospins, are
subject to these constraints. The exchange energy (\ref{AppenSSBb}) may be
used as the effective Hamiltonian for the Goldstone modes. However, there
are $N^{2}-1$ real components in $\mathcal{S}_{a}(\mathbf{x})$, but not all
of them are independent. Because of this fact, except for the case of SU(2),
it is quite difficult to analyze Goldstone modes based on (\ref{AppenSSBb}).
Later we shall derive the effective Hamiltonian in terms of the CP$^{N-1}$
field: See (\ref{HamilNCCP}).

\section{Quantum Hall Skyrmions}

We proceed to analyze topological excitations. As we have shown, a
topological soliton necessarily carries the electron number. The simplest
charge excitation is the hole excitation. It is curious but, as we shall
soon see, a hole is a kind of skyrmion in the noncommutative plane. A
generic skyrmion state is constructed as a W$_{\infty }$(N)-rotated state of
the hole state. We start with a skyrmion with $Q=1$, and subsequently pass
to a multi-skyrmion with $Q=k$. It is to be remarked that the Hamiltonian is
not necessary to develop a theory of skyrmions. What we use is only the fact
that the ground state is a spontaneously symmetry broken one.

\subsection{Microscopic Skyrmions}

The ground state is expressed as

\end{subequations}
\begin{equation}
|\text{g}\rangle =\prod\limits_{n=0}^{\infty }g_{\mu }c_{\mu }^{\dag
}(n)|0\rangle ,  \label{GrounState}
\end{equation}%
where a constant vector $g_{\mu }$ stands for the isospin component
spontaneously chosen, obeying the normalization condition%
\begin{equation}
\bar{g}_{\mu }g_{\mu }=1.  \label{StepCPa}
\end{equation}%
The ground state satisfies 
\begin{subequations}
\begin{align}
\rho (m,n)|\text{g}\rangle =& \delta _{mn}|\text{g}\rangle , \\
\langle \text{g}|S_{a}(m,n)|\text{g}\rangle =& \frac{1}{2}\delta _{mn}\bar{g}%
_{\mu }(\lambda _{a})_{\mu \nu }g_{\nu }.
\end{align}%
The simplest charge excitation is a hole excitation given by 
\end{subequations}
\begin{equation}
|\text{h}\rangle =\bar{g}_{\mu }c_{\mu }(0)|\text{g}\rangle ,
\label{HoleExcit}
\end{equation}%
with its energy $\langle $h$|H_{\text{V}}|$h$\rangle =\epsilon _{\text{X}}$.

We consider a W$_{\infty }$(N)-rotated state of the hole state $|$h$\rangle $%
, 
\begin{equation}
|\mathfrak{S}_{\text{sky}}\rangle =e^{iW}|\text{h}\rangle
=e^{iW}\prod\limits_{n=0}^{\infty }g_{\mu }c_{\mu }^{\dag }(n+1)|0\rangle ,
\label{SkyrmFormuA}
\end{equation}%
where $W$ is an element of the W$_{\infty }$(N) algebra (\ref{WalgebP}). As
we have noticed in (\ref{ElectNumbeA}), the electron number of this state is
the same as that of the hole state, or $\Delta N_{\text{e}}^{\text{cl}}=-1$.

The simplest W$_{\infty }$(N) rotation mixes only the nearest neighboring
sites, and is given by the choice of $W=\sum_{n=0}^{\infty }W_{n}$ with 
\begin{equation}
iW_{n}=\alpha _{n}h_{\mu }g_{\nu }^{\ast }c_{\mu }^{\dag }(n)c_{\nu
}(n+1)-\alpha _{n}h_{\mu }^{\ast }g_{\nu }c_{\nu }^{\dag }(n+1)c_{\mu }(n),
\end{equation}%
where $\alpha _{n}$ is a real parameter, and $h_{\mu }$ is a constant vector
orthogonal to $g_{\mu }$,%
\begin{equation}
\bar{h}_{\mu }h_{\mu }=1,\quad \bar{g}_{\mu }h_{\mu }=\bar{h}_{\mu }g_{\mu
}=0.  \label{StepCPb}
\end{equation}%
Note that $W_{n}$ is a Hermitian operator belonging to the W$_{\infty }$(N)
algebra. Remark that $[W_{n},W_{m}]=0$, as follows from (\ref{StepCPb}) in
the case of $m=n+1$, and trivially in all other cases. We define%
\begin{equation}
\xi ^{\dagger }(n)=e^{+iW}g_{\mu }c_{\mu }^{\dag }(n+1)e^{-iW}.
\label{WtransA}
\end{equation}%
It is easy to show that 
\begin{equation}
\xi ^{\dagger }(n)=u_{\mu }(n)c_{\mu }^{\dag }(n)+v_{\mu }(n)c_{\mu }^{\dag
}(n+1),  \label{SkyrmCreatOpera}
\end{equation}%
where we have set 
\begin{equation}
u_{\mu }(n)=h_{\mu }\sin \alpha _{n},\quad v_{\mu }(n)=g_{\mu }\cos \alpha
_{n}.
\end{equation}%
The conditions 
\begin{subequations}
\label{CondiOnUV}
\begin{align}
\bar{u}_{\mu }(n)u_{\mu }(n)+\bar{v}_{\mu }(n)v_{\mu }(n)=& 1,
\label{CondiOnU} \\
\bar{v}_{\mu }(n)u_{\mu }(n+1)=& 0,  \label{CondiOnV}
\end{align}%
follow on $u_{\mu }(n)$ and $v_{\mu }(n)$. It is necessary that 
\end{subequations}
\begin{equation}
\lim_{n\rightarrow \infty }u_{\mu }(n)=0,\qquad \lim_{n\rightarrow \infty
}v_{\mu }(n)=g_{\mu },  \label{CondiUVasymp}
\end{equation}%
since the skyrmion state should approach the ground state asymptotically.

The operator $\xi (m)$ satisfies the standard canonical anticommutation
relation,%
\begin{equation}
\{\xi (m),\xi ^{\dagger }(n)\}=\delta _{mn},\quad \{\xi (m),\xi (n)\}=0,
\label{AntiCommu}
\end{equation}%
as is verified with the use of the condition (\ref{CondiOnUV}). The hole
state is a special limit of the skyrmion state with $u_{\mu }(n)=0$ and $%
v_{\mu }(n)=g_{\mu }$ for all $n$.

It is now easy to see that the W$_{\infty }$(N)-rotated state (\ref%
{SkyrmFormuA}) is rewritten as%
\begin{equation}
|\mathfrak{S}_{\text{sky}}\rangle =\prod\limits_{n=0}^{\infty }\xi ^{\dagger
}(n)|0\rangle  \label{MicroSkyrmState}
\end{equation}%
by the use of (\ref{WtransA}) and $e^{-iW_{n}}|0\rangle =0$. This agrees
with the skyrmion state\cite{Fertig94B} proposed to carry out a Hartree-Fock
approximation.

Multi-skyrmion states are given by (\ref{MicroSkyrmState}) with 
\begin{equation}
\xi ^{\dag }(n)=u_{\mu }(n)c_{\mu }^{\dag }(n)+v_{\mu }(n)c_{\mu }^{\dag
}(n+k),  \label{SkyrmCreat}
\end{equation}%
where $k$ indicates the amount of skyrmions. Normalization is still given by
(\ref{CondiOnU}) but (\ref{CondiOnV}) replaced by $\bar{v}_{\mu }(n)u_{\mu
}(n+k)=0$. In a single skyrmion case the nearest neighboring sites are mixed
in operators $\xi (n)$. For $k=2$ the mixing appears among the sites with
even $n$ and separately among the ones with odd $n$, while no mixing appears
among even and odd sites. In this sense there arise two separate chains of
mixed sites, where the above steps can be carried out separately. In the
case of general $k$ there are $k$ separate chains, so nothing essentially
new occurs, and the analog of (\ref{SkyrmFormuA}) appears as%
\begin{equation}
|\mathfrak{S}_{\text{sky}}\rangle =e^{iW}\prod_{n=0}^{\infty }g_{\mu }c_{\mu
}^{\dag }(n+k)|0\rangle =e^{iW}\prod_{n=0}^{k-1}\bar{g}_{\mu }c_{\mu }(n)|%
\text{g}\rangle ,  \label{MultiSkyrm}
\end{equation}%
where $|$g$\rangle $ is the ground state given by (\ref{GrounState}). The
electron number is evidently given by $\Delta N_{\text{e}}^{\text{cl}}=-k$.

\subsection{Noncommutative CP$^{N-1}$ Skyrmions}

We construct the noncommutative CP$^{N-1}$ field describing the
multi-skyrmion state (\ref{MicroSkyrmState}) with (\ref{SkyrmCreat}). For
this purpose we calculate the quantities 
\begin{equation}
D_{\mu \nu }^{\text{cl}}(m,n)=\langle \mathfrak{S}_{\text{sky}}|c_{\nu
}^{\dag }(n)c_{\mu }(m)|\mathfrak{S}_{\text{sky}}\rangle  \label{StepCPd}
\end{equation}%
for the multi-skyrmion state. It is straightforward to show%
\begin{equation}
c_{\mu }(m)|\mathfrak{S}_{\text{sky}}\rangle =\left[ u_{\mu }(m)\xi
(m)+v_{\mu }(m-\alpha )\xi (m-\alpha )\right] |\mathfrak{S}_{\text{sky}%
}\rangle ,  \label{StepCPc}
\end{equation}%
where the convention $v_{\mu }(m)=0$ for $m<0$ is introduced. We then use 
\begin{equation}
\langle \mathfrak{S}_{\text{sky}}|\xi ^{\dag }(m)\xi (n)|\mathfrak{S}_{\text{%
sky}}\rangle =\delta _{mn}
\end{equation}%
together with (\ref{StepCPc}), to find the only nonvanishing values of $%
D_{\mu \nu }^{\text{cl}}(m,n)$ to be 
\begin{align}
D_{\mu \nu }^{\text{cl}}(n,n)=& u_{\mu }(n)\bar{u}_{\nu }(n)+v_{\mu }(n-k)%
\bar{v}_{\nu }(n-k),  \notag \\
D_{\mu \nu }^{\text{cl}}(n,n+k)=& u_{\mu }(n)\bar{v}_{\nu }(n),  \notag \\
D_{\mu \nu }^{\text{cl}}(n+k,n)=& v_{\mu }(n)\bar{u}_{\nu }(n).
\label{StepCPe}
\end{align}%
This gives the Weyl-ordered operator as%
\begin{align}
& 2\pi \theta \,O[D_{\mu \nu }^{\text{cl}}]  \notag \\
=& \left[ u_{\mu }(n)\bar{u}_{\nu }(n)+v_{\mu }(n-k)\bar{v}_{\nu }(n-k)%
\right] |n\rangle \langle n|  \notag \\
& +u_{\mu }(n)\bar{v}_{\nu }(n)|n\rangle \langle n+k|+v_{\mu }(n)\bar{u}%
_{\nu }(n)|n+k\rangle \langle n|.
\end{align}%
Comparing this with (\ref{DinCPw}) we uniquely come to%
\begin{equation}
\mathfrak{n}_{\mu }=\sum_{n=0}^{\infty }\left[ u_{\mu }(n)|n\rangle \langle
n|+v_{\mu }(n)|n+k\rangle \langle n|\right] ,  \label{CPforSkyrm}
\end{equation}%
whose symbol is\cite{Harvey}%
\begin{align}
n_{\mu }(\mathbf{x})=& 2e^{-\bar{z}z}\sum_{n=0}^{\infty }(-1)^{n}u_{\mu
}(n)L_{n}(2\bar{z}z)  \notag \\
& +2^{\frac{k}{2}+1}z^{k}e^{-\bar{z}z}\sum_{n=0}^{\infty }\frac{(-1)^{n}%
\sqrt{n!}}{\sqrt{(n+\alpha )!}}v_{\mu }(n)L_{n}^{k}(2\bar{z}z).
\label{MultiSkyrmX}
\end{align}%
This is the noncommutative CP$^{N-1}$ field describing the multi-skyrmion
state.

\subsection{Topological Charge of Skyrmions}

We have already shown that the topological charge is given by $Q=-\Delta N_{%
\text{e}}^{\text{cl}}=k$ for the multi-skyrmion state via the charge-number
relation (\ref{RelatChargNumbe}). We confirm this by an explicit calculation.

In the Weyl form the topological charge (\ref{TopolChargNC}) is expressed as%
\begin{equation}
Q=\text{Tr}([\mathfrak{n}_{\mu }^{\dag },\mathfrak{n}_{\mu }]).
\end{equation}%
Here, we cannot use Tr$([A,B])=$Tr$(AB)-$Tr$(BA)=0$, which is valid only if
Tr$(AB)$ and Tr$(BA)$ are separately well defined. This is not the case
here, and the trace operation must be carried out after the commutator is
calculated.

Using (\ref{CPforSkyrm}) we obtain%
\begin{equation}
\lbrack \mathfrak{n}_{\mu }^{\dag },\mathfrak{n}_{\mu }]=\sum_{n=0}^{\infty }%
\bar{v}_{\mu }(n)v_{\mu }(n)\left[ |n\rangle \langle n|-|n+k\rangle \langle
n+k|\right] .
\end{equation}%
This is still inappropriate for calculating the trace, since the separate
pieces are divergent due to the behavior $|v_{\mu }(n)|\rightarrow 1$ as $%
n\rightarrow \infty $, as implied by the boundary condition (\ref%
{CondiUVasymp}). We rewrite the formula in terms of $u_{\mu }(n)$ instead of 
$v_{\mu }(n)$, and obtain 
\begin{align}
\lbrack \mathfrak{n}_{\mu }^{\dag },\mathfrak{n}_{\mu }]=&
\sum_{n=0}^{k-1}|n\rangle \langle n|-\sum_{n=0}^{\infty }\bar{u}_{\mu
}(n)u_{\mu }(n)|n\rangle \langle n|  \notag \\
& +\sum_{n=0}^{\infty }\bar{u}_{\mu }(n)u_{\mu }(n)|n+k\rangle \langle n+k|.
\label{StepCPf}
\end{align}%
Each piece is well defined since $u_{\mu }(n)\rightarrow 0$ as $n\rightarrow
\infty $, and the traces can be calculated separately, which eventually
leads to $Q=k$.

\subsection{Hole State}

In order to understand the difference of the noncommutative theory from the
commutative one, it is instructive to analyze the hole state (\ref{HoleExcit}%
) more in details. To make the argument as simple as possible, we consider
the SU(2) spin system with the ground state being the up-spin filled one,%
\begin{equation}
|\text{g}\rangle =\prod\limits_{n=0}^{\infty }c_{\uparrow }^{\dag
}(n)|0\rangle .
\end{equation}%
The hole state is%
\begin{equation}
|\text{h}\rangle =c_{\uparrow }(0)|\text{g}\rangle ,
\end{equation}%
with its bare density being 
\begin{subequations}
\label{BareDensiHole}
\begin{align}
\hat{\rho}^{\text{cl}}(\mathbf{x})=& \rho _{0}\left( 1-2e^{-r^{2}/\theta
}\right) ,  \label{BareDensiHoleA} \\
\hat{S}_{x}^{\text{cl}}(\mathbf{x})=& \hat{S}_{y}^{\text{cl}}(\mathbf{x}%
)=0,\quad \hat{S}_{z}^{\text{cl}}(\mathbf{x})=\frac{1}{2}\hat{\rho}^{\text{cl%
}}(\mathbf{x}).  \label{BareDensiHoleB}
\end{align}%
It represents literally a hole of the size $\sqrt{\theta }$. The spin
texture is trivial.

In the commutative theory, the topological number is given by (\ref{TopolCPy}%
), or 
\end{subequations}
\begin{equation}
Q=\frac{1}{2\pi i}\epsilon _{kl}\sum_{s}\int \!d^{2}x\,(\partial _{k}\bar{n}%
_{\mu })(\partial _{l}n_{\mu }),  \label{TopolCPz}
\end{equation}%
which is equivalent to the Pontryagin number,%
\begin{equation}
Q_{\text{P}}={\frac{1}{\pi }}\int \!d^{2}x\,\varepsilon _{abc}\varepsilon
_{ij}\mathcal{S}_{a}(\partial _{i}\mathcal{S}_{b})(\partial _{j}\mathcal{S}%
_{c}).  \label{PontrNumbe}
\end{equation}%
Here $\mathcal{S}_{a}$ is the normalized spin density, $\mathcal{S}%
_{a}\equiv \frac{1}{2}\bar{n}_{\mu }(\sigma _{a})_{\mu \nu }n_{\nu }$, with $%
\sigma _{a}$ the Pauli matrix. We would obviously conclude $Q_{\text{P}}=0$
for the trivial spin texture such as (\ref{BareDensiHoleB}).

However, this is not the case in the noncommutative theory. According to the
argument presented in the previous subsection, its topological charge is
given by%
\begin{equation}
Q=\text{Tr}([\mathfrak{n}_{\mu }^{\dag },\mathfrak{n}_{\mu }])=\text{Tr}%
(|0\rangle \langle 0|)=1,  \label{TopolChargK}
\end{equation}%
since $u_{\mu }(n)=0$ for all $n$ in (\ref{StepCPf}). There is no mystery
here. Let us explain this by calculating the topological number explicitly
in the coordinate space.

The CP$^{1}$ field which gives the trivial spin texture (\ref{BareDensiHoleB}%
) via (\ref{RhoSinN}) is highly nontrivial, 
\begin{subequations}
\begin{align}
n_{\uparrow }(\mathbf{x})=& 2^{\frac{3}{2}}ze^{-\bar{z}z}\sum_{n=0}^{\infty }%
\frac{(-1)^{n}}{\sqrt{n+1}}L_{n}^{1}(2\bar{z}z),  \label{StepCPg} \\
n_{\downarrow }(\mathbf{x})=& 0.
\end{align}%
It is well defined everywhere. Only the asymptotic behavior contributes to
the topological number (\ref{TopolChargNC}) in the noncommutative
formulation, or equivalently to (\ref{TopolCPz}) in the commutative
formulation. Any finite order terms in $n$ vanish in the limit $\bar{z}%
z\rightarrow \infty $ due to the term $e^{-\bar{z}z}$ in (\ref{StepCPg}).
Thus, in calculating the topological charge, we may replace $1/\sqrt{(n+1)}$
by $\Gamma (n+\frac{3}{2})/\Gamma (n+2)$, which is valid for $n\gg 1$. We
then use \cite{Hansen} 
\end{subequations}
\begin{equation}
\sum_{n=0}^{\infty }\frac{(a)_{n}}{(c)_{n}}t^{n}L_{n}^{(c-1)}(x)=\frac{1}{%
(1-t)^{a}}M\left( a;c;\frac{xt}{t-1}\right) ,
\end{equation}%
where $M(a;b;x)$ is the Kummer function. Taking $a=\frac{3}{2}$, $c=2$ and $%
t=-1$, we obtain%
\begin{equation}
n_{\uparrow }(\mathbf{x})\rightarrow ze^{-\bar{z}z}\frac{\Gamma (\frac{3}{2})%
}{\Gamma (2)}M\left( \frac{3}{2};2;\bar{z}z\right) \rightarrow \frac{z}{|z|}%
=e^{i\vartheta },
\end{equation}%
where $\vartheta $ is the azimuthal angle. It carries a proper winding
number. The topological number of a hole is clearly $Q=1$, as is consistent
with (\ref{TopolChargK}).

Hence we conclude that a hole is a kind of skyrmion in the noncommutative
theory: Its spin texture is trivial but the associated CP$^{N-1}$ field is
highly nontrivial and carries a proper winding number.

It is easy to see what happens in the commutative limit for the hole state,
where the CP$^{1}$ field (\ref{StepCPg}) is reduced to%
\begin{equation}
n_{\uparrow }(\mathbf{x})=e^{i\vartheta },\qquad n_{\downarrow }(\mathbf{x}%
)=0.
\end{equation}%
It is ill defined at the origin. Hence we cannot calculate the topological
charge by the formula (\ref{TopolCPz}) for the hole state. It is the
essential feature of the noncommutative theory that such a singularity is
regulated over the region of area $\theta $. This is the reason why the
noncommutative CP$^{1}$ field for the hole state is well defined everywhere
and carries a winding number.

\section{Hard-Core Interaction}

We have so far analyzed the kinematical structure of multi-skyrmions in the
noncommutative plane. The CP$^{N-1}$ field given by (\ref{MultiSkyrmX})
contains infinitely many parameters $u_{\mu }(n)$ and $v_{\mu }(n)$. These
variables are to be fixed so as to minimize the energy of the state. Let us
carry out this program by taking the hard-core interaction\cite{LeeD01} for
the potential $V\left( \mathbf{x}\right) $ in the Hamiltonian (\ref%
{CouloHamil}), 
\begin{equation}
V(\mathbf{x}-\mathbf{y})=\delta ^{2}(\mathbf{x}-\mathbf{y}).
\label{ContaPoten}
\end{equation}%
The Hamiltonian reads%
\begin{equation}
H_{\text{hc}}={\frac{1}{2}}\int \!d^{2}x\;\Delta \rho (\mathbf{x})\Delta
\rho (\mathbf{x}).  \label{ContaHamil}
\end{equation}%
For simplicity we restrict the isospin SU(N) symmetry to the spin SU(2)
symmetry in what follows.

The matrix element (\ref{Vmnij}) is easily calculable to give%
\begin{equation}
V_{mnij}=\frac{1}{8\pi \theta }\frac{\sqrt{(m+i)!}}{\sqrt{m!i!}}\frac{\sqrt{%
(n+j)!}}{\sqrt{n!j!}}\frac{\delta _{m+i,n+j}}{\sqrt{2^{m+i+n+j}}},
\label{VmnijHard}
\end{equation}%
which possesses the additional symmetry separately with respect to $%
m\leftrightarrow i$ and $n\leftrightarrow j$. Due to this symmetry the
Hamiltonian (\ref{ContaHamil}) is reduced to 
\begin{equation}
H_{\text{hc}}=2\sum_{mnij}V_{mnij}c_{\uparrow }^{\dag }(m)c_{\downarrow
}^{\dag }(i)c_{\downarrow }(j)c_{\uparrow }(n)-\frac{1}{4\pi \theta }\Delta
N_{\text{e}}^{\text{cl}},
\end{equation}%
or%
\begin{equation}
H_{\text{hc}}=\int \!d^{2}x\;\psi _{\uparrow }^{\dagger }(\mathbf{x})\psi
_{\downarrow }^{\dagger }(\mathbf{x})\psi _{\downarrow }(\mathbf{x})\psi
_{\uparrow }(\mathbf{x})-\frac{1}{4\pi \theta }\Delta N_{\text{e}}^{\text{cl}%
}.  \label{ContaHamilx}
\end{equation}%
where we have used 
\begin{equation}
\epsilon _{\text{D}}=\epsilon _{\text{X}}=\frac{1}{4\pi \theta },
\end{equation}%
which follows from (\ref{EnergD}) and (\ref{EnergX}).

\subsection{Factorizable Skyrmions}

We consider the multi-skyrmion state (\ref{MicroSkyrmState}) with (\ref%
{SkyrmCreat}), or%
\begin{equation}
\xi ^{\dag }(n)=u(n)c_{\downarrow }^{\dag }(n)+v(n)c_{\uparrow }^{\dag
}(n+k).
\end{equation}%
Due to the anticommutation relation (\ref{AntiCommu}) we have%
\begin{equation}
\langle \mathfrak{S}_{\text{sky}}|\xi ^{\dag }(m)\xi ^{\dag }(i)\xi (j)\xi
(n)|\mathfrak{S}_{\text{sky}}\rangle =\delta _{mn}\delta _{ij}-\delta
_{mj}\delta _{in},
\end{equation}%
and using (\ref{VmnijHard}) we obtain%
\begin{align}
& \langle \mathfrak{S}_{\text{sky}}|H_{\text{hc}}|\mathfrak{S}_{\text{sky}%
}\rangle =\frac{k}{4\pi \theta }+\frac{1}{8\pi \theta }\sum_{mn}\frac{%
(m+n+k)!}{2^{m+n+k}m!n!}  \notag \\
& \times \left\vert \frac{\sqrt{m!}}{\sqrt{(m+k)!}}v(m)u(n)-\frac{\sqrt{n!}}{%
\sqrt{(n+k)!}}v(n)u(m)\right\vert ^{2},
\end{align}%
where we have accounted $\Delta N_{\text{e}}^{\text{cl}}=-k$. The second
term, being positive semidefinite, takes the minimum for%
\begin{equation}
\frac{\sqrt{n!}}{\sqrt{(n+k)!}}\frac{v(n)}{u(n)}=\frac{1}{\omega ^{k}},
\label{StepCPh}
\end{equation}%
where $\omega $ is an arbitrary complex constant. This gives 
\begin{subequations}
\label{ExpliUV}
\begin{align}
|u(n)|^{2}& =\frac{|\omega |^{2k}}{|\omega |^{2k}+[(n+1)\cdots (n+k)]}, \\
|v(n)|^{2}& =\frac{(n+1)\cdots (n+k)}{|\omega |^{2k}+[(n+1)\cdots (n+k)]}
\end{align}%
with the aid of the condition $|u(n)|^{2}+|v(n)|^{2}=1$.

Using (\ref{StepCPc}) we can actually verify that 
\end{subequations}
\begin{equation}
H_{\text{hc}}|\mathfrak{S}_{\text{sky}}\rangle =-\frac{1}{4\pi \theta }%
\Delta N^{\text{cl}}|\mathfrak{S}_{\text{sky}}\rangle =\frac{k}{4\pi \theta }%
|\mathfrak{S}_{\text{sky}}\rangle ,  \label{ContaSkyrmEnerg}
\end{equation}
when parameters $u_{\mu }(n)$ and $v_{\mu }(n)$ are given by (\ref{ExpliUV}%
). Consequently, $|\mathfrak{S}_{\text{sky}}\rangle $ is an eigenstate of
the Hamiltonian (\ref{ContaHamilx}) with $\Delta N^{\text{cl}}=-k$. The
eigenvalue is independent of the scale parameter $\omega $, and the skyrmion
state is degenerate with the hole state.

The CP$^{1}$ field (\ref{CPforSkyrm}) corresponding to the multi-skyrmion
state is $|\mathfrak{S}_{\text{sky}}\rangle $ expressed as 
\begin{subequations}
\begin{align}
\mathfrak{n}_{\uparrow } =&\sum_{n}v(n)|n+k\rangle \langle n|,
\label{StepCPx} \\
\mathfrak{n}_{\downarrow } =&\sum_{n}u(n)|n\rangle \langle n|
\end{align}%
with (\ref{ExpliUV}). We calculate this explicitly with the aid of (\ref%
{ExpliUV}).

From (\ref{FockState}) we have 
\end{subequations}
\begin{equation}
(b^{\dag })^{k}|n\rangle =\left[ \frac{(n+k)!}{n!}\right] ^{\frac{1}{2}%
}|n+k\rangle ,
\end{equation}%
which leads to%
\begin{equation}
(b^{\dag })^{k}\mathfrak{n}_{\downarrow }=\sum_{n=0}^{\infty }u(n)\left[ 
\frac{(n+k)!}{n!}\right] ^{\frac{1}{2}}|n+k\rangle \langle n|.
\end{equation}%
Here we use (\ref{StepCPh}) and (\ref{StepCPx}) to find%
\begin{equation}
(b^{\dag })^{k}\mathfrak{n}_{\downarrow }=\omega ^{k}\mathfrak{n}_{\uparrow
},
\end{equation}%
which in terms of symbols turns into%
\begin{equation}
z^{k}\star n_{\downarrow }(\mathbf{x})=(\sqrt{2}\omega )^{k}n_{\uparrow }(%
\mathbf{x}),
\end{equation}%
where $z\star z^{k}=z^{k+1}$ has been used.

Then the CP field reads%
\begin{equation}
n(\mathbf{x})=\frac{1}{\lambda ^{k}}\left[ 
\begin{array}{c}
z^{k} \\ 
\lambda ^{k}%
\end{array}%
\right] \star n_{\downarrow }(\mathbf{x}),  \label{NCCP1}
\end{equation}%
with $\lambda \equiv \sqrt{2}\omega $, where $n_{\downarrow }(\mathbf{x})$
is the function of $r=|\mathbf{x}|$ only%
\begin{equation}
n_{\downarrow }(\mathbf{x})=2e^{-\bar{z}z}\sum_{n=0}^{\infty
}(-1)^{n}u(n)L_{n}(2\bar{z}z).
\end{equation}%
It is to be noticed that the spin part is factorized as in (\ref{NCCP1}). It
is reminiscent of the skyrmion CP$^{1}$ field%
\begin{equation}
n(\mathbf{x})=\frac{1}{\sqrt{|z|^{2k}+|\lambda |^{2k}}}\left[ 
\begin{array}{c}
z^{k} \\ 
\lambda ^{k}%
\end{array}%
\right]  \label{CP1Skyrm}
\end{equation}%
in the ordinary CP$^{1}$ model (\ref{ModelCP}).

Finally we calculate the physical quantities $\rho ^{\text{cl}}(\mathbf{x})$
and $S_{a}^{\text{cl}}(\mathbf{x})$ for the multi-skymion state, 
\begin{subequations}
\begin{align}
\rho ^{\text{cl}}(\mathbf{x})& =\langle \mathfrak{S}_{\text{sky}}|\psi
^{\dag }(\mathbf{x})\psi (\mathbf{x})|\mathfrak{S}_{\text{sky}}\rangle , \\
S_{a}^{\text{cl}}(\mathbf{x})& =\frac{1}{2}\langle \mathfrak{S}_{\text{sky}%
}|\psi ^{\dag }(\mathbf{x})\sigma _{a}\psi (\mathbf{x})|\mathfrak{S}_{\text{%
sky}}\rangle .
\end{align}%
Substituting (\ref{StepCPe}) with (\ref{ExpliUV}) into (\ref{ProjeDensiClass}%
), we obtain $\hat{D}_{\mu \nu }^{\text{cl}}(\mathbf{k})$. Then, we
construct $D_{\mu \nu }^{\text{cl}}(\mathbf{x})$ from (\ref{DensiDy}), which
may be decomposed into $\rho ^{\text{cl}}(\mathbf{x})$ and $S_{a}^{\text{cl}%
}(\mathbf{x})$.

In this way we come to 
\end{subequations}
\begin{subequations}
\label{DensiSkyrm}
\begin{align}
\rho ^{\text{cl}}(\mathbf{x})& =\frac{1}{2\pi \theta }\left[ \left( \frac{%
r^{2}}{2\theta }\right) ^{k}+|\omega |^{2k}\right] f\left( \frac{r^{2}}{%
2\theta }\right) ,  \label{StepSkyrmA} \\
S_{x}^{\text{cl}}(\mathbf{x})& =\frac{+1}{2\pi \theta }\left( \frac{|\omega
|r}{\sqrt{2\theta }}\right) ^{k}f\left( \frac{r^{2}}{2\theta }\right) \cos
k\vartheta , \\
S_{y}^{\text{cl}}(\mathbf{x})& =\frac{-1}{2\pi \theta }\left( \frac{|\omega
|r}{\sqrt{2\theta }}\right) ^{k}f\left( \frac{r^{2}}{2\theta }\right) \sin
k\vartheta , \\
S_{z}^{\text{cl}}(\mathbf{x})& =\frac{1}{4\pi \theta }\left[ \left( \frac{%
r^{2}}{2\theta }\right) ^{k}-|\omega |^{2k}\right] f\left( \frac{r^{2}}{%
2\theta }\right) ,
\end{align}%
where $f(x)$ is given by 
\end{subequations}
\begin{equation}
f(x)=e^{-x}\sum_{n=0}^{\infty }\frac{1}{|\omega |^{2k}+[(n+1)\cdots (n+k)]}%
\frac{x^{n}}{n!}  \label{StepSkyrmB}
\end{equation}%
and $\vartheta $ is the azimuthal angle. It follows from (\ref{DensiSkyrm})
that%
\begin{equation}
S_{a}^{\text{cl}}(\mathbf{x})=\rho ^{\text{cl}}(\mathbf{x})\mathcal{S}_{a}(%
\mathbf{x}),  \label{SkyrmConfiConta}
\end{equation}%
where%
\begin{equation}
\mathcal{S}_{a}(\mathbf{x})=\frac{1}{2}\bar{n}(\mathbf{x})\sigma _{a}n(%
\mathbf{x})
\end{equation}%
with (\ref{CP1Skyrm}). It is the nonlinear spin field normalized as $%
\mathcal{S}_{a}(\mathbf{x})\mathcal{S}_{a}(\mathbf{x})=1/4$, and given by 
\begin{subequations}
\label{FactoSkyrmS}
\begin{align}
\mathcal{S}_{x}(\mathbf{x})=& \frac{\left( |\lambda |r\sqrt{\theta }\right)
^{k}}{\left( r^{2}\right) ^{k}+\left( |\lambda |^{2}\theta \right) ^{k}}\cos
k\vartheta , \\
\mathcal{S}_{y}(\mathbf{x})=& \frac{-\left( |\lambda |r\sqrt{\theta }\right)
^{k}}{\left( r^{2}\right) ^{k}+\left( |\lambda |^{2}\theta \right) ^{k}}\sin
k\vartheta , \\
\mathcal{S}_{z}(\mathbf{x})=& \frac{1}{2}\frac{\left( r^{2}\right)
^{k}-\left( |\lambda |^{2}\theta \right) ^{k}}{\left( r^{2}\right)
^{k}+\left( |\lambda |^{2}\theta \right) ^{k}}.
\end{align}%
It agrees with the skyrmion spin field in the ordinary O(3) nonlinear sigma
model. The factorizability (\ref{SkyrmConfiConta}) together with (\ref%
{FactoSkyrmS}) is a peculiar property. We call such a skyrmion a
factorizable skyrmion.

\subsection{Zeeman interaction}

QH effects occur in the external magnetic field. The electron couples with
the magnetic field via the Zeeman term in the realistic system, 
\end{subequations}
\begin{equation}
H_{\text{Z}}=-\Delta _{\text{Z}}\int \!d^{2}x\,S_{z}(\mathbf{x}),
\label{SpinZeema}
\end{equation}%
where $\Delta _{Z}=|g|\mu _{B}B_{\perp }$ is the Zeeman gap with $\mu _{B}$
the Bohr magneton and $g$ the magnetic g-factor. It is interesting to
calculate the Zeeman energy of the factorizable skyrmion (\ref%
{SkyrmConfiConta}).

For simplicity we study the simplest skyrmion with $Q=k=1$. We may rewrite (%
\ref{StepSkyrmB}) as 
\begin{equation}
f(x)=\frac{1}{|\omega |^{2}+1}e^{-x}M(|\omega |^{2}+1;|\omega |^{2}+2;x).
\end{equation}%
It behaves as%
\begin{equation}
\lim_{x\rightarrow \infty }f(x)=\frac{1}{x}-\frac{|\omega |^{2}}{x^{2}}+%
\frac{|\omega |^{2}(|\omega |^{2}-1)}{x^{3}}.
\end{equation}%
Hence we obtain 
\begin{subequations}
\begin{align}
\lim_{r\rightarrow \infty }\rho ^{\text{cl}}(\mathbf{x}) =&\frac{1}{2\pi
\theta }\left( 1-\frac{2\theta ^{2}|\lambda |^{2}}{r^{4}}\right) , \\
\lim_{r\rightarrow \infty }S_{z}^{\text{cl}}(\mathbf{x}) =&\frac{1}{4\pi
\theta }\left( 1-\frac{2\theta |\lambda |^{2}}{r^{2}}\right) .
\label{StepSkyrmD}
\end{align}%
The number of spins flipped around a skyrmion is given by 
\end{subequations}
\begin{equation}
N_{\text{spin}}=\int \!d^{2}x\;\left\{ S_{z}^{\text{cl}}(\mathbf{x})-\frac{1%
}{4\pi \theta }\right\} ,
\end{equation}%
which is identified with the skyrmion spin. We find $N_{\text{spin}}$ to
diverge logarithmically, unless $\lambda =0$, due to the asymptotic behavior
(\ref{StepSkyrmD}). The Zeeman energy $\langle H_{\text{Z}}\rangle =-\Delta
_{\text{Z}}N_{\text{spin}}$ is divergent, except for the hole, from the
infrared contribution however small the Zeeman effect is.

The factorizable skyrmion (\ref{SkyrmConfiConta}) is no longer valid in the
hard-core model with the Zeeman term. There exists surely a skyrmion state
which has a finite Zeeman energy once the factorizability is abandoned: See (%
\ref{CondiOnUVb}) for an example. Nevertheless, we can show that the hole
state has the lowest energy. The reason reads as follows. The factorizable
skyrmion is an eigenstate of the hard-core Hamiltonian, $H_{\text{hc}}|%
\mathfrak{S}_{\text{sky}}\rangle =E_{\text{hc}}|\mathfrak{S}_{\text{sky}%
}\rangle $ with $E_{\text{hc}}=|\Delta N^{\text{cl}}|/4\pi \theta $, as in (%
\ref{ContaSkyrmEnerg}). Accordingly any spin texture $|\mathfrak{S}\rangle $
possessing the same electron number $\Delta N^{\text{cl}}$ has a higher
energy, $\langle \mathfrak{S}|H_{\text{hc}}|\mathfrak{S}\rangle \geq E_{%
\text{hc}}$. Furthermore its Zeeman energy is larger than that of the hole, $%
\langle \mathfrak{S}|H_{\text{Z}}|\mathfrak{S}\rangle \geq \frac{1}{2}\Delta
_{\text{Z}}$. Hence, 
\begin{equation}
\langle \mathfrak{S}|\left( H_{\text{hc}}+H_{\text{Z}}\right) |\mathfrak{S}%
\rangle \geq E_{\text{hc}}+\frac{1}{2}\Delta _{\text{Z}},
\end{equation}%
where the equality holds for the hole state. Consequently there are no
skyrmions in the presence of the Zeeman interaction in the system with the
hard-core interaction.

\section{Noncommutative CP$^{N-1}$ Model}

In the previous section we have studied the skyrmion state (\ref{CPforSkyrm}%
) in the hard-core model (\ref{ContaPoten}) and determined parameters $%
u_{\mu }(n)$ and $v_{\mu }(n)$ explicitly. We have emphasized that the
skyrmion itself is independent of the Hamiltonian. For the sake of
completeness we analyze the skyrmion state (\ref{CPforSkyrm}) in the
noncommutative CP$^{N-1}$ model.

The noncommutative CP$^{N-1}$ model is defined by the Hamiltonian\cite%
{LeePLB01}%
\begin{equation}
H_{\text{CP}}=\kappa \theta ^{2}\text{Tr}\left[ (\partial _{m}\mathfrak{n}%
_{\mu }^{\dag })(\partial _{m}\mathfrak{n}_{\mu })+(\mathfrak{n}_{\mu
}^{\dag }\partial _{m}\mathfrak{n}_{\mu })(\mathfrak{n}_{\nu }^{\dag
}\partial _{m}\mathfrak{n}_{\nu })\right] ,
\end{equation}%
where the derivative is given by (\ref{DerivNCs}) or%
\begin{equation}
\partial _{i}\mathfrak{n}_{\mu }=-\frac{i}{\theta }\epsilon _{ij}\left[
X_{j},\mathfrak{n}_{\mu }\right] .
\end{equation}%
It is convenient to introduce the notation 
\begin{align}
\mathcal{D}_{z}\mathfrak{n}_{\mu }& \equiv \mathcal{D}_{x}\mathfrak{n}_{\mu
}-i\mathcal{D}_{y}\mathfrak{n}_{\mu },  \notag \\
\mathcal{D}_{\bar{z}}\mathfrak{n}_{\mu }& \equiv \mathcal{D}_{x}n_{\mu }+i%
\mathcal{D}_{y}\mathfrak{n}_{\mu },
\end{align}%
where%
\begin{equation}
\mathcal{D}_{i}\mathfrak{n}_{\mu }\equiv -i\theta ^{-1}\epsilon _{ij}\left(
[X_{j},\mathfrak{n}_{\mu }]-\mathfrak{n}_{\mu }\mathfrak{n}_{\gamma }[X_{j},%
\mathfrak{n}_{\gamma }]\right) ,
\end{equation}%
and rewrite the noncommutative CP$^{N-1}$ model (\ref{HamilNCCP}) as%
\begin{equation}
E_{\text{X}}=\frac{1}{2}\kappa \theta ^{2}\text{Tr}\left[ (\mathcal{D}_{z}%
\mathfrak{n}_{\mu })^{\dag }(\mathcal{D}_{z}\mathfrak{n}_{\mu })+(\mathcal{D}%
_{\bar{z}}\mathfrak{n}_{\mu })^{\dag }(\mathcal{D}_{\bar{z}}\mathfrak{n}%
_{\mu })\right] .  \label{HamilNCCPx}
\end{equation}%
We have already defined the topological charge $Q$ by the formula (\ref%
{TopolChargNC}). However, there are other equivalent ways of defining it,
and here we use the representation 
\begin{equation}
Q=\frac{1}{2}\theta \text{Tr}\left[ (\mathcal{D}_{z}\mathfrak{n}_{\mu
})^{\dag }(\mathcal{D}_{z}\mathfrak{n}_{\mu })-(\mathcal{D}_{\bar{z}}%
\mathfrak{n}_{\mu })^{\dag }(\mathcal{D}_{\bar{z}}\mathfrak{n}_{\mu })\right]
.  \label{TopolCPx}
\end{equation}%
The equivalence is readily proved because the topological charge densities
associated with (\ref{TopolChargNC}) and (\ref{TopolCPx}) are different only
by a total derivative term that does not contribute to the topological
charge.

The BPS energy bound follows from (\ref{HamilNCCPx}) and (\ref{TopolCPx}),%
\begin{equation}
E_{\text{X}}\geq \kappa \theta |Q|.  \label{BogomBound}
\end{equation}%
This bound becomes saturated for 
\begin{equation}
\mathcal{D}_{\bar{z}}\mathfrak{n}_{\mu }=0\hspace*{5mm}\hspace*{5mm}\text{%
(skyrmions).}  \label{BogomSkyrm}
\end{equation}%
or%
\begin{equation}
\mathcal{D}_{z}\mathfrak{n}_{\mu }=0\hspace*{5mm}\hspace*{5mm}\text{%
(antiskyrmions).}
\end{equation}%
We consider the case of skyrmions in what follows. The self-dual equation (%
\ref{BogomSkyrm}) takes the form 
\begin{equation}
\lbrack b^{\dag },\mathfrak{n}_{\mu }]-\mathfrak{n}_{\mu }\mathfrak{n}%
_{\gamma }[b^{\dag },\mathfrak{n}_{\gamma }]=0.  \label{BogomEqA}
\end{equation}%
We consider explicitly the case of SU(2) and look for the solution to (\ref%
{BogomEqA}) in the form (\ref{CPforSkyrm}), or 
\begin{equation}
\mathfrak{n}_{\uparrow }=\sum_{m=0}^{\infty }v(m)|m+k\rangle \langle
m|,\quad \mathfrak{n}_{\downarrow }=\sum_{m=0}^{\infty }u(m)|m\rangle
\langle m|,
\end{equation}%
where $|u(k)|^{2}+|v(k)|^{2}=1$. This describes a multi-skyrmion carrying
the topological charge $Q=k$. We get 
\begin{align}
\lbrack b^{\dag },\mathfrak{n}_{\uparrow }]-\mathfrak{n}_{\uparrow }%
\mathfrak{n}_{\mu }[b^{\dag },\mathfrak{n}_{\mu }]& =\sum_{m=0}^{\infty
}f_{\uparrow }(m)|m+k+1\rangle \langle m|,  \notag \\
\lbrack b^{\dag },\mathfrak{n}_{\downarrow }]-\mathfrak{n}_{\downarrow }%
\mathfrak{n}_{\mu }[b^{\dag },\mathfrak{n}_{\mu }]& =\sum_{m=0}^{\infty
}f_{\downarrow }(m)|m+1\rangle \langle m|,  \label{BogomEqBd}
\end{align}%
with 
\begin{align}
f_{\uparrow }(m)=& v(m)\bar{u}(m+1)u(m+1)\sqrt{m+k+1}  \notag \\
& -v(m+1)\bar{u}(m+1)u(m)\sqrt{m+1},  \notag \\
f_{\downarrow }(m)=& u(m)\bar{v}(m+1)v(m+1)\sqrt{m+1}  \notag \\
& -u(m+1)\bar{v}(m+1)v(m)\sqrt{m+k+1}.
\end{align}%
The equation (\ref{BogomEqA}) holds if $f_{\uparrow }(m)=0$ and $%
f_{\downarrow }(m)=0$. They lead to a single equation, which can be
summarized as 
\begin{equation}
\frac{v(m)}{u(m)}=\frac{\sqrt{m+k}}{\sqrt{m}}\frac{v(m-1)}{u(m-1)}.
\end{equation}%
This recurrence relation is solved as 
\begin{equation}
\frac{v(m)}{u(m)}=\frac{\sqrt{(m+k)!}}{\sqrt{m!}}\frac{1}{\omega ^{k}},
\label{StepCPi}
\end{equation}%
where we have introduced the complex parameter $\omega $ by 
\begin{equation}
\omega ^{k}\equiv \frac{u(0)}{v(0)}k!.
\end{equation}%
The parameters (\ref{StepCPi}) are found to be the same as (\ref{StepCPh}).

Consequently the skyrmions are identical both in the hard-core model and the
noncommutative CP$^{N-1}$ model, though these two models are very different.
We discuss the relation between them in Section \ref{SecDiscus}.

\section{Effective Theory}

We have explored kinematical properties of the classical density matrix $%
\hat{D}_{\mu \nu }^{\text{cl}}$. It is interesting to discuss the dynamical
property of the classical density, especially, what the equation of motion
for $\hat{D}_{\mu \nu }^{\text{cl}}$ is.

\subsection{Classical Equation of Motion}

We start with the quantum equation of motion for the density matrix ${D}%
_{\mu \nu }(k,l)$,

\begin{equation}
i\hbar \frac{d}{dt}D_{\mu \nu }(k,l)=[{D}_{\mu \nu }(k,l),H_{\text{V}}],
\label{EquatOfD}
\end{equation}%
where $H_{\text{V}}$ is the four-fermion interaction Hamiltonian (\ref%
{CouloHamil}). By using (\ref{WalgebD}) it is explicitly calculated as 
\begin{align}
i\hbar \frac{d}{dt}D_{\gamma \alpha }(k,l)=& V_{mnkj}{D}_{\mu \mu }(n,m){D}%
_{\gamma \alpha }(j,l)  \notag \\
& -V_{mnil}{D}_{\mu \mu }(n,m){D}_{\gamma \alpha }(k,i)  \notag \\
& +V_{knij}{D}_{\gamma \alpha }(n,l){D}_{\mu \mu }(j,i)  \notag \\
& -V_{mlij}{D}_{\gamma \alpha }(k,m){D}_{\mu \mu }(j,i).
\end{align}%
We take the average of this equation with respect to the Fock state (\ref%
{SkyrmFormuC}). Using

\begin{align}
\langle \mathfrak{S}|{D}_{\nu \mu }& (n,m){D}_{\tau \sigma }(j,i)|\mathfrak{%
S\rangle }  \notag \\
=& {D}_{\nu \mu }^{\text{cl}}(n,m){D}_{\tau \sigma }^{\text{cl}}(j,i)-{D}%
_{\tau \mu }^{\text{cl}}(j,m){D}_{\nu \sigma }^{\text{cl}}(n,i)  \notag \\
& +\delta _{\nu \sigma }\delta _{ni}{D}_{\tau \mu }^{\text{cl}}(j,m),
\end{align}%
where ${D}_{\nu \mu }^{\text{cl}}(n,m)=\langle \mathfrak{S}|{D}_{\nu \mu
}(n,m)|\mathfrak{S\rangle }$, we obtain%
\begin{align}
i\hbar \frac{d}{dt}D_{\gamma \alpha }^{\text{cl}}(l,k)=& 2V_{mnlj}{D}_{\mu
\mu }^{\text{cl}}(n,m){D}_{\gamma \alpha }^{\text{cl}}(j,k)  \notag \\
& -2V_{mnik}{D}_{\mu \mu }^{\text{cl}}(n,m){D}_{\gamma \alpha }^{\text{cl}%
}(l,i)  \notag \\
& -2V_{mnlj}{D}_{\gamma \mu }^{\text{cl}}(j,m){D}_{\mu \alpha }^{\text{cl}%
}(n,k)  \notag \\
& +2V_{mnik}{D}_{\gamma \mu }^{\text{cl}}(l,m){D}_{\mu \alpha }^{\text{cl}%
}(n,i).  \label{StepDensiA}
\end{align}%
Provided the classical density ${D}_{\nu \mu }^{\text{cl}}(n,m)$ is endowed
with the Poisson structure%
\begin{align}
i\hbar \lbrack & {D}_{\mu \nu }^{\text{cl}}(m,n),{D}_{\sigma \tau }^{\text{cl%
}}(i,j)]_{\text{PB}}  \notag \\
& =\delta _{\mu \tau }\delta _{mj}{D}_{\sigma \nu }^{\text{cl}}(i,n)-\delta
_{\sigma \nu }\delta _{in}{D}_{\mu \tau }^{\text{cl}}(m,j),  \label{PoissDmn}
\end{align}%
it is straightforward to show that (\ref{StepDensiA}) is summarized into%
\begin{equation}
\frac{d}{dt}D_{\mu \nu }^{\text{cl}}(m,n)=[{D}_{\mu \nu }^{\text{cl}%
}(m,n),E_{\text{V}}]_{\text{PB}},  \label{EquatOfDx}
\end{equation}%
where $E_{\text{V}}\equiv \langle \mathfrak{S}|H_{\text{V}}|\mathfrak{S}%
\rangle $ is the average of the Hamiltonian. As we have noticed in (\ref%
{DecomFormu}), it consists of the direct energy (\ref{EnergDa}) and the
exchange energy (\ref{EnergXa}).

It is remarkable that the Poisson structure (\ref{PoissDmn}) is precisely
the same as the algebra (\ref{WalgebD}). To see its significance more in
details we combine it with the magnetic-translation group property (\ref%
{MagneTrans}),%
\begin{align}
2\pi i\hbar \lbrack \hat{D}_{\mu \nu }^{\text{cl}}(\mathbf{k}),\hat{D}%
_{\sigma \tau }^{\text{cl}}(\mathbf{k}^{\prime })]_{\text{PB}}=& \delta
_{\mu \tau }e^{+\frac{i}{2}\theta \mathbf{k}\wedge \mathbf{k}^{\prime }}\hat{%
D}_{\sigma \nu }^{\text{cl}}(\mathbf{k}+\mathbf{k}^{\prime })  \notag \\
& -\delta _{\sigma \nu }e^{-\frac{i}{2}\theta \mathbf{k}\wedge \mathbf{k}%
^{\prime }}\hat{D}_{\mu \tau }^{\text{cl}}(\mathbf{k}+\mathbf{k}^{\prime }).
\label{StepNCc}
\end{align}%
It corresponds to the W$_{\infty }$(N) algebra (\ref{WalgebP}), indicating
that the classical density $\hat{D}_{\mu \nu }^{\text{cl}}$ should obey the W%
$_{\infty }$(N) algebra as well.

It is important that the Poisson structure (\ref{StepNCc}) is resolved in
terms of the noncommutative CP$^{N-1}$ field. The relation (\ref{DinCPone})
reads%
\begin{equation}
\hat{D}_{\mu \nu }^{\text{cl}}(\mathbf{k})=\frac{1}{4\pi ^{2}\theta }\int
\!d^{2}k^{\prime }\,e^{-\frac{i}{2}\theta \mathbf{k}\wedge \mathbf{k}%
^{\prime }}n_{\mu }(\mathbf{k}^{\prime })\bar{n}_{\nu }(\mathbf{k}^{\prime }-%
\mathbf{k})
\end{equation}%
in the momentum space, where $\bar{n}_{\mu }(\mathbf{k})$ is the complex
conjugate of $n_{\mu }(\mathbf{k})$. Substituting this into (\ref{StepNCc}),
provided%
\begin{equation}
i\hbar \lbrack n_{\mu }(\mathbf{k}),\bar{n}_{\nu }(\mathbf{k}^{\prime })]_{%
\text{PB}}=2\pi \theta \delta _{\mu \nu }(\mathbf{k}^{\prime }-\mathbf{k}),
\end{equation}%
or%
\begin{equation}
i\hbar \lbrack n_{\mu }(\mathbf{x}),\bar{n}_{\nu }(\mathbf{y})]_{\text{PB}%
}=2\pi \theta \delta _{\mu \nu }(\mathbf{x}-\mathbf{y}),
\label{PoissBrackNCCP}
\end{equation}%
we can easily show that (\ref{StepNCc}) holds as an identity.

\subsection{Derivative Expansion}

We proceed to derive the low energy effective theory of the Hamiltonian
system (\ref{CouloHamil}) in terms of the CP$^{N-1}$ field. The
corresponding energy consists of the direct and exchange energies $E_{\text{D%
}}$ and $E_{\text{X}}$ as given by (\ref{EnergDa}) and (\ref{EnergXa}),
respectively. The direct energy is not interesting since it represents
simply the classical energy of a charge distribution $-e\Delta \rho ^{\text{%
cl}}(\mathbf{x})$.

We analyze the exchange energy $E_{\text{X}}$. Corresponding to (\ref{DinCPw}%
) we find%
\begin{equation}
D_{\mu \nu }^{\text{cl}}(m,n)=\langle m|\mathfrak{n}_{\mu }\mathfrak{n}_{\nu
}^{\dag }|n\rangle .
\end{equation}%
Substituting this into (\ref{EnergXa}) we come to 
\begin{equation}
E_{\text{X}}=-\int \frac{d^{2}k}{4\pi }e^{-\frac{1}{2}\theta k^{2}}V(\mathbf{%
k})\text{Tr}\left( \mathfrak{n}_{\mu }\mathfrak{n}_{\nu }^{\dag }e^{i\mathbf{%
kX}}\mathfrak{n}_{\nu }\mathfrak{n}_{\mu }^{\dag }e^{-i\mathbf{kX}}\right)
+N_{\text{e}}^{\text{cl}}\epsilon _{\text{X}},
\end{equation}%
where we have taken (\ref{Vmnij}) into account. To bring it to a more
reasonable form, we use 
\begin{equation}
N_{\text{e}}^{\text{cl}}=\text{Tr}\left( \mathfrak{n}_{\mu }\mathfrak{n}%
_{\mu }^{\dag }\right) ,
\end{equation}%
and rewrite the exchange energy as 
\begin{align}
E_{\text{X}}=& \int \frac{d^{2}k}{4\pi }e^{-\frac{1}{2}\theta k^{2}}V(%
\mathbf{k})  \notag \\
& \times \text{Tr}\left( \mathfrak{n}_{\mu }\mathfrak{n}_{\mu }^{\dag }-%
\mathfrak{n}_{\mu }\mathfrak{n}_{\nu }^{\dag }e^{i\mathbf{kX}}\mathfrak{n}%
_{\nu }\mathfrak{n}_{\mu }^{\dag }e^{-i\mathbf{kX}}\right) ,
\end{align}%
where the cancellation of divergences occurs at $\mathbf{k}=0$. We may
change the order of operators under the trace simultaneously in both terms
so that the cancellation at $\mathbf{k}=0$ is maintained. We achieve at%
\begin{equation}
E_{\text{X}}=\int \frac{d^{2}k}{4\pi }e^{-\frac{1}{2}\theta k^{2}}V(\mathbf{k%
})\text{Tr}\left( \mathbb{I}-\mathfrak{n}_{\mu }^{\dag }e^{i\mathbf{kX}}%
\mathfrak{n}_{\mu }\cdot \mathfrak{n}_{\nu }^{\dag }e^{-i\mathbf{kX}}%
\mathfrak{n}_{\nu }\right)  \label{ExchaNCCP}
\end{equation}%
where we have used the normalization $\mathfrak{n}_{\mu }^{\dag }\mathfrak{n}%
_{\mu }=\mathbb{I}$.

As far as we are concerned about sufficient smooth field configurations we
may make the derivative expansion and take the lowest order term. Here we
note that%
\begin{align}
& \mathbb{I}-\mathfrak{n}_{\mu }^{\dag }e^{i\mathbf{kX}}\mathfrak{n}_{\mu
}\cdot \mathfrak{n}_{\nu }^{\dag }e^{-i\mathbf{kX}}\mathfrak{n}_{\nu } 
\notag \\
=& \mathbb{[}\mathfrak{n}_{\mu }^{\dag },e^{i\mathbf{kX}}][e^{-i\mathbf{kX}},%
\mathfrak{n}_{\mu }]-\mathfrak{n}_{\mu }^{\dag }[e^{i\mathbf{kX}},\mathfrak{n%
}_{\mu }]\mathfrak{n}_{\nu }^{\dag }[e^{-i\mathbf{kX}},\mathfrak{n}_{\nu }].
\end{align}%
Because the derivative is given by (\ref{DerivNCs}) or%
\begin{equation}
\partial _{i}\mathfrak{n}_{\mu }=-\frac{i}{\theta }\epsilon _{ij}\left[
X_{j},\mathfrak{n}_{\mu }\right] ,
\end{equation}%
the derivative expansion corresponds to the expansion in $\mathbf{X}$,%
\begin{equation}
e^{i\mathbf{kX}}=\mathbb{I}+ik_{i}X_{i}-\frac{1}{2}%
k_{i}k_{j}X_{i}X_{j}+O(X^{3}).
\end{equation}%
Substituting this into (\ref{ExchaNCCP}) and assuming the rotational
invariance of $V(\mathbf{k})$ we integrate over the angle in the momentum
space. In such a way, up to the lowest order term of the derivative
expansion, we come to 
\begin{align}
E_{\text{X}}=& \kappa \text{Tr}\left( \mathfrak{n}_{\mu }^{\dag }X_{i}X_{i}%
\mathfrak{n}_{\mu }-\mathfrak{n}_{\mu }^{\dag }X_{i}\mathfrak{n}_{\mu }%
\mathfrak{n}_{\nu }^{\dag }X_{i}\mathfrak{n}_{\nu }\right)  \notag \\
=& \kappa \text{Tr}\left( [\mathfrak{n}_{\mu }^{\dag },X_{i}][X_{i},%
\mathfrak{n}_{\mu }]-\mathfrak{n}_{\mu }^{\dag }[X_{i},\mathfrak{n}_{\mu }]%
\mathfrak{n}_{\nu }^{\dag }[X_{i},\mathfrak{n}_{\nu }]\right)  \notag \\
=& \kappa \theta ^{2}\text{Tr}\left[ (\partial _{m}\mathfrak{n}_{\mu }^{\dag
})(\partial _{m}\mathfrak{n}_{\mu })+(\mathfrak{n}_{\mu }^{\dag }\partial
_{m}\mathfrak{n}_{\mu })(\mathfrak{n}_{\nu }^{\dag }\partial _{m}\mathfrak{n}%
_{\nu })\right] ,  \label{HamilNCCP}
\end{align}%
where the constant $\kappa $ is given by 
\begin{equation}
\kappa \equiv \frac{1}{4}\int dk\,k^{3}e^{-\frac{1}{2}\theta k^{2}}V(k).
\label{KappaV}
\end{equation}%
It agrees with the noncommutative CP$^{N-1}$ model\cite{LeePLB01}. It is
interesting that the noncommutative CP$^{N-1}$ model is derived from the
four-fermion interaction Hamiltonian irrespective of the explicit form of
the potential $V\left( \mathbf{x}\right) $ as the lowest order term in the
derivative expansion.

We can use the noncommutative CP$^{N-1}$ model (\ref{HamilNCCP}) as the
effective Hamiltonian for the Goldstone modes, which are small fluctuations
of the CP$^{N-1}$ field around the ground state. Such fluctuations occur
without the density excitation ($\Delta \rho _{\text{e}}^{\text{cl}}(\mathbf{%
x})=0$), thus minimizing the direct energy as $E_{\text{D}}=0$. The
condition reads 
\begin{equation}
n_{\mu }(\mathbf{x})\star \bar{n}_{\mu }(\mathbf{x})=\bar{n}_{\mu }(\mathbf{x%
})\star n_{\mu }(\mathbf{x})=1.
\end{equation}%
Furthermore, sufficiently smooth isospin waves are described in its
commutative limit,%
\begin{equation}
E_{\text{X}}=\kappa \theta ^{2}\int d^{2}x\left[ (\partial _{m}n_{\mu
}^{\dag })(\partial _{m}n_{\mu })+(n_{\mu }^{\dag }\partial _{m}n_{\mu
})(n_{\nu }^{\dag }\partial _{m}n_{\nu })\right] ,  \label{ModelCP}
\end{equation}%
which is the ordinary CP$^{N-1}$ model. There are $N-1$ complex Goldstone
modes.

\section{Discussion}

\label{SecDiscus}

We have analyzed the kinematical and dynamical properties of the classical
density matrix $\hat{D}_{\mu \nu }^{\text{cl}}(\mathbf{x})$ in the
noncommutative plane. In particular we have elucidated noncommutative
solitons by taking a concrete instance of the QH system with the SU(N)
symmetry.

A microscopic skyrmion state with $Q=1$ is a W$_{\infty }$(N)-rotated state
of a hole state. The skyrmion texture is given by (\ref{GenerSkyrm}), or%
\begin{equation}
\mathfrak{n}_{\mu }=\sum_{n=0}^{\infty }\left[ u_{\mu }(n)|n\rangle \langle
n|+v_{\mu }(n)|n+1\rangle \langle n|\right]
\end{equation}%
with $|u(n)|^{2}+|v(n)|^{2}=1$. It contains infinitely many parameters. They
are fixed when the Hamiltonian is given. We have studied explicitly the
hard-core model and the noncommutative CP$^{N-1}$ model. In these models all
those parameters are determined analytically except for the scale parameter $%
\omega $,%
\begin{equation}
u^{2}(n)=\frac{\omega ^{2}}{n+1+\omega ^{2}}.  \label{ParamU}
\end{equation}%
The skyrmion energy is independent of the parameter $\omega $.

It is intriguing that the skyrmions are identical both in the hard-core
model and the noncommutative CP$^{N-1}$ model. Let us examine this
equivalence more in details. On one hand, in the noncommutative CP$^{N-1}$
model the skyrmion energy is given by 
\begin{equation}
E_{\text{CP}}=\kappa \theta k,  \label{EnergNCCPmodel}
\end{equation}%
as follows from (\ref{BogomBound}). On the other hand, in the hard-core
model it is given by (\ref{ContaSkyrmEnerg}), or%
\begin{equation}
E_{\text{hc}}=\frac{k}{4\pi \theta }.
\end{equation}%
The parameter $\kappa $ is given by (\ref{KappaV}), which reads%
\begin{equation}
\kappa =\frac{1}{4\pi \theta ^{2}}.
\end{equation}%
Hence, these two energies are identical, $E_{\text{CP}}=E_{\text{hc}}$.
Namely, in the hard-core model the skyrmion energy coming from the
lowest-order term saturates the total energy. It implies that the direct
energy and the higher order exchange energy have precisely cancelled out
each other. Such a cancellation is possible since both of them are short
range in the hard-core model.

The skyrmion solution obtained in these models satisfies the factorizability
property%
\begin{equation}
S_{a}^{\text{cl}}(\mathbf{x})=\rho ^{\text{cl}}(\mathbf{x})\mathcal{S}_{a}(%
\mathbf{x}),
\end{equation}%
where $\mathcal{S}_{a}(\mathbf{x})=\frac{1}{2}\bar{n}(\mathbf{x})\lambda
_{a}n(\mathbf{x})$ with the Gell-mann matrix $\lambda _{a}$. We have called
them factorizable skyrmions. It is this class of skyrmions that has been
studied in almost all previous works on QH systems. For instance we have
studied skyrmions\cite{Ezawa99L,EzawaX03B} in the semiclassical
approximation based on the composite-boson picture, but it allows us only to
consider such a limited class of excitations.

However, as we have shown, the energy of the factorizable skyrmion is
infinite once the Zeeman effect is taken into account. Consequently
realistic skyrmions in the QH system are not factorizable skyrmions. The
factorizability follows from the particular form (\ref{ParamU}) of
parameters $u(n)$. Any other choice produces an unfactorizable skyrmion.

The electron-electron interaction is given by the Coulomb potential $V(%
\mathbf{x})=e^{2}/4\pi \varepsilon |\mathbf{x}|$ in the realistic QH system.
Since the interaction decreases continuously as the distance increases, the
direct Coulomb energy $E_{\text{D}}$ decreases as the skyrmion size
increases. However, a large skyrmion requires a large Zeeman energy. Hence,
when the Zeeman effect is very large, what is excited is the hole state (\ref%
{HoleExcit}). Thus the total Coulomb energy is%
\begin{equation}
E_{\text{C}}=\epsilon _{\text{X}}=\sqrt{\frac{\pi }{2}}\frac{e^{2}}{8\pi
\varepsilon \ell _{B}},\qquad \text{(hole)}.
\end{equation}%
On the other hand, when the Zeeman effect is infinitesimal, an infinitely
large skyrmion is excited. It is well approximated by the factorizable
skyrmion, where the total Coulomb energy is%
\begin{equation}
E_{\text{C}}=\frac{1}{2}\epsilon _{\text{X}},\qquad \text{(large skyrmion)}.
\end{equation}%
It is one half of the excitation energy of one hole.

To determine a skyrmion excitation in general, it is necessary to carry out
a numerical calculation by making an appropriate anzats on parameters $u(n)$%
. For instance it is reasonable to choose%
\begin{equation}
u^{2}(n)=\frac{\omega ^{2}t^{2n+2}}{n+1+\omega ^{2}},  \label{CondiOnUVb}
\end{equation}%
since the parameter $t$ interpolates smoothly the hole ($t=0$) and the
factorizable skyrmion ($t=1$). As is reported elsewhere\cite{Tsitsishvili05}%
, it is possible to explain the experimental data\cite{Schmeller95L} based
on this anzats in all range of the Zeeman gap $\Delta _{\text{Z}}$.

One of our main results is the rigorous relation between the electron number
density and the topological charge density, 
\begin{equation}
\Delta \rho ^{\text{cl}}(\mathbf{x})=-J_{0}(\mathbf{x})=-\frac{1}{2\pi
\theta }\sum_{\mu }[\bar{n}_{\mu }(\mathbf{x}),n_{\mu }(\mathbf{x})]_{\star
}.
\end{equation}%
The topological charge density thus defined enjoys a privileged role that it
is a physical quantity essentially equal to the electron density excitation
induced by a topological soliton. Accordingly the noncommutative soliton
carries necessarily the electric charge. The topological charge $Q=\int
\!d^{2}x\,J_{0}(\mathbf{x})$ becomes an integer for a wide class of states (%
\ref{SkyrmFormuC}). This class of states contains all physically relevant
states at integer filling factors. However, it does not contain states
relevant at fractional filling factors. In such a system the topological
charge $Q$ would become fractional since it carries a fractional electric
charge\cite{Laughlin83L}. We wish to generalize the present scheme to
fractional QH systems in a future work.

\section{Acknowledgments}

We are grateful to the hospitality of Theoretical Physics Laboratory, RIKEN,
where a part of this work was done. One of the authors (ZFE) is supported in
part by Grants-in-Aid for Scientific Research from Ministry of Education,
Science, Sports and Culture (Nos. 13135202,14540237). The other author (GT)
acknowledges a research fellowship from Japan Society for Promotion of
Science (Nos. L04514).

\appendix

\section{Decomposition Formula}

\label{AppenA}

The basic element in our analysis is the density matrix 
\begin{equation}
D_{\mu \nu }^{\text{cl}}(m,n)=\langle \mathfrak{S}|c_{\nu }^{\dag }(n)c_{\mu
}(m)|\mathfrak{S}\rangle,  \label{DensiMatri}
\end{equation}%
obeying the relation (\ref{DDD}), or 
\begin{equation}
\sum_{s}D_{\mu \sigma }^{\text{cl}}(m,s)D_{\sigma \nu }^{\text{cl}%
}(s,n)=D_{\mu \nu }^{\text{cl}}(m,n).  \label{DDDx}
\end{equation}%
The decomposition formula (\ref{DecomFormu}) on the classical energy, or%
\begin{equation}
E_{\text{V}}\equiv \langle \mathfrak{S}|H_{\text{V}}|\mathfrak{S}\rangle =E_{%
\text{D}}+E_{\text{X}}  \label{DecomFormuApp}
\end{equation}%
with (\ref{EnergDX}), has played an important role. We prove these relations.

We consider a system with a finite number of Landau sites $(m=0,1,\ldots
,N_{\Phi }-1)$ assuming the limit $N_{\Phi }\rightarrow \infty $ in final
expressions. It is important that the W$_{\infty }$(N) algebra (\ref{WalgebD}%
) is not violated in such a finite system. For the sake of convenience we
combine the isospin and site indices into a multi-index $M\equiv (\mu ,m)$,
which runs over the values $M=1,2,\ldots ,NN_{\Phi }$. In terms of
multi-indices the W$_{\infty }$(N) algebra turns out to be an algebra U(NN$%
_{\Phi }$), and the transformation rules for fermion operators are 
\begin{subequations}
\label{A2}
\begin{align}
e^{-iW}c_{I}e^{+iW}& =\sum_{I^{\prime }}(U)_{II^{\prime }}c_{I^{\prime }}, \\
e^{-iW}c_{I}^{\dag }e^{+iW}& =\sum_{I^{\prime }}c_{I^{\prime }}^{\dag
}(U^{\dag })_{I^{\prime }I},
\end{align}%
where $U$ is an $(NN_{\Phi })\times (NN_{\Phi })$ unitary matrix 
\end{subequations}
\begin{equation}
UU^{\dag }=U^{\dag }U=\mathbb{I}_{(NN_{\Phi })\times (NN_{\Phi })}.
\label{A3}
\end{equation}%
The state $|\mathfrak{S}_{0}\rangle $ given by (\ref{SkyrmFormuCx}) is
expressed as%
\begin{equation}
|\mathfrak{S}_{0}\rangle =\prod_{K=1}^{NN_{\Phi }}\left[ c_{K}^{\dag }\right]
^{\nu _{K}}|0\rangle .  \label{A4}
\end{equation}%
Now it is easy to see%
\begin{equation}
\langle \mathfrak{S}_{0}|c_{I}^{\dag }c_{J}|\mathfrak{S}_{0}\rangle =\nu
_{J}\delta _{IJ},  \label{A5}
\end{equation}%
where $\delta _{MN}\equiv \delta _{\mu \nu }\delta _{mn}$.

The density matrix (\ref{DensiMatri}) appears as%
\begin{equation}
D_{IJ}^{\text{cl}}=\langle \mathfrak{S}|c_{J}^{\dag }c_{I}|\mathfrak{S}%
\rangle .
\end{equation}%
Employing $|\mathfrak{S}\rangle =e^{iW}|\mathfrak{S}_{0}\rangle $ together
with (\ref{A2}) and (\ref{A5}) we come to%
\begin{equation}
D_{IJ}^{\text{cl}}=\sum_{K}\nu _{K}(U)_{IK}(U^{\dag })_{KJ}.
\end{equation}%
From this we get%
\begin{equation}
\sum_{J}D_{IJ}^{\text{cl}}D_{JK}^{\text{cl}}=\sum_{J}(\nu
_{J})^{2}(U)_{IJ}(U^{\dag })_{JK},
\end{equation}%
where we have used (\ref{A3}).

Since $\nu _{J}=0$ or $1$, we have $(\nu _{J})^{2}=\nu _{J}$ and come to%
\begin{equation}
\sum_{J}D_{IJ}^{\text{cl}}D_{JK}^{\text{cl}}=\sum_{J}\nu
_{J}(U)_{IJ}(U^{\dag })_{JK}=D_{IK}^{\text{cl}},
\end{equation}%
which is nothing but (\ref{DDDx}), or (\ref{DDD}).

We next calculate the quantity $\langle \mathfrak{S}|c_{I}^{\dag
}c_{J}^{\dag }c_{K}c_{L}|\mathfrak{S}\rangle $. For this purpose we use the
same technique and also%
\begin{equation}
\langle \mathfrak{S}_{0}|c_{I}^{\dag }c_{J}^{\dag }c_{K}c_{L}|\mathfrak{S}%
_{0}\rangle =\nu _{K}\nu _{L}(\delta _{JK}\delta _{IL}-\delta _{IK}\delta
_{JL}),
\end{equation}%
which holds due to the given particular structure of (\ref{A4}).

Collecting everything together we come to%
\begin{equation}
\langle \mathfrak{S}|c_{I}^{\dag }c_{J}^{\dag }c_{K}c_{L}|\mathfrak{S}%
\rangle =D_{KJ}^{\text{cl}}D_{LI}^{\text{cl}}-D_{KI}^{\text{cl}}D_{LJ}^{%
\text{cl}},
\end{equation}%
which turns out to yield the decomposition formula (\ref{DecomFormuApp}), or
(\ref{DecomFormu}).


\begin{thebibliography}{99}
\bibitem{BookConnes} A. Connes, Noncommutative Geometry (Academic Press,
1994).

\bibitem{Nekrasov98} N. Nekrasov and A. Schwarz, Commun. Math. Phys. 198
(1998) 689.

\bibitem{Gopakumar00} R. Gopakumar, S. Minwalla and A. Strominger, JHEP 0005
(2000) 020.

\bibitem{Schaposnik04} F. Schaposnik, Braz. J. Phys. \textbf{34} (2004) 1349.

\bibitem{LeePLB01} B.-H. Lee, K. Lee and H.S. Yang, Phys. Lett. \textbf{B498}
(2001) 277.

\bibitem{groenewold} H. Groenewold, Physica \textbf{12} (1946) 405;.

\bibitem{moyal} J. Moyal, Proc. Camb. Phil. Soc. \textbf{45} (1949) 99.

\bibitem{BookEzawa} Z.F. Ezawa, \textit{Quantum Hall Effects: Field
Theoretical Approach and Related Topics} (World Scientific, 2000).

\bibitem{BookDasSarma} S. Das Sarma and A. Pinczuk (eds), \textit{%
Perspectives in Quantum Hall Effects} (Wiley, 1997).

\bibitem{Girvin84B} S.M. Girvin and T. Jach, Phys. Rev. B29 (1984) 5617.

\bibitem{Girvin85L} S.M. Girvin, A.H. MacDonald and P.M. Platzman, Phys.
Rev. Lett. 54 (1985) 581.

\bibitem{Iso92PLB} S. Iso, D. Karabali, and B. Sakita, Phys. Lett. B \textbf{%
196}, 143 (1992).

\bibitem{Cappelli93NPB} A. Cappelli, C. Trugenberger, and G. Zemba, Nucl.
Phys. B \textbf{396} (1993) 465.

\bibitem{Moon95B} K. Moon, H. Mori, K. Yang, S.M. Girvin, A.H. MacDonald, L.
Zheng, D. Yoshioka and S-C. Zhang, Phys. Rev. B 51 (1995) 5138.

\bibitem{Ezawa97B} Z.F. Ezawa, Phys. Rev. B \textbf{55} (1997) 7771.

\bibitem{Sondhi93B} S.L. Sondhi, A. Karlhede, S.A. Kivelson and E.H. Rezayi,
Phys. Rev. B \textbf{47} (1993) 16419.

\bibitem{Barrett95L} S.E. Barrett, G. Dabbagh, L.N. Pfeiffer, K.W. West and
R. Tycko, Phys. Rev. Lett. \textbf{74} (1995) 5112.

\bibitem{Aifer96L} E.H. Aifer, B.B. Goldberg and D.A. Broido, Phys. Rev.
Lett. \textbf{76 } (1996) 680.

\bibitem{Schmeller95L} A. Schmeller, J.P. Eisenstein, L.N. Pfeiffer and K.W.
West, Phys. Rev. Lett. \textbf{75} (1995) 4290.

\bibitem{Fertig94B} H.A. Fertig, L. Brey, R. Cote and A.H. MacDonald, Phys.
Rev. B 50 (1994) 11018.

\bibitem{Abolfath97B} M. Abolfath, J.J. Palacios, H.A. Fertig, S.M. Girvin
and A.H. MacDonald, Phys. Rev. \textbf{B} 56 (1997) 6795.

\bibitem{MacDonald96B} A.H. MacDonald, H.A. Fertig and L. Brey, Phys. Rev.
Lett. 76 (1996) 2153.

\bibitem{Weyl} H. Weyl, Z. Physik 46 (1927) 1.

\bibitem{EzawaX03B} Z.F. Ezawa, G. Tsitsishvili and K. Hasebe, Phys. Rev.
B67 (2003) 125314.

\bibitem{Ezawa99L} Z.F. Ezawa, Phys. Rev. Lett. \textbf{82} (1999) 3512.

\bibitem{LeeD01} B.-H. Lee, K. Moon and C. Rim, Phys. Rev. \textbf{D64}
(2001) 085014.

\bibitem{Harvey} J.A. Harvey, \textit{Komaba Lectures on Noncommutative
Solitons and D-Branes}, hep-th/0102076.

\bibitem{macfarlain} A.J. MacFarlane, Phys. Lett. \textbf{B82} (1979) 239.

\bibitem{Hansen} E.R. Hansen, \textit{A table of series and products}
(Prentice-Hall, 1975); see (48.7.7) in this book.

\bibitem{Tsitsishvili05} G. Tsitsishvili and Z.F. Ezawa, to be published in
Phys. Rev. B.

\bibitem{Laughlin83L} R.B. Laughlin, Phys. Rev. Lett. 50 (1983) 1395.
\end{thebibliography}
\end{document}